\documentclass[fleqn,usenatbib]{mnras}

\usepackage[T1]{fontenc}
\usepackage{ae,aecompl}

\usepackage{graphicx}
\usepackage{amsmath}
\usepackage{amssymb}
\usepackage{amsfonts}
\usepackage{natbib}
\usepackage{graphicx}
\usepackage{epsfig}
\include{journaldefs}
\usepackage[caption=false]{subfig}
\usepackage{url}
\usepackage{textcomp}
\usepackage{float}
\usepackage[T1]{fontenc} 
\usepackage{aecompl}
\hypersetup{draft}
\def\fig{Figure}

\def\sect{Sect.}

\def\tab{Table}

\def\Tab{Table}

\def\arcm{$^{\prime}\,$}
\def\arcs{$^{\prime\prime}\,$}
\def\mujybm   {${\rm \mu}$Jy\,beam$^{-1}$}
\def\mjybm   {${\rm m}$Jy\,beam$^{-1}$}

\title{Jets, Arcs and Shocks: NGC\,5195 at radio wavelengths}

\author[H.~Rampadarath et al.]{H.~Rampadarath$^{1}$\thanks{E-mail: haydenrampadarath@gmail.com}, R. Soria$^{2,3,4}$\thanks{E-mail: Roberto.Soria@curtin.edu.au}, R.~Urquhart$^{3}$, M.~K.~Argo$^{5,1}$, M.~Brightman$^{6,7}$, 
\newauthor
C.~K.~Lacey$^{8}$, E.~M.~Schlegel$^{9}$, R.~J.~Beswick$^{1}$,  R.~D.~Baldi$^{10}$, T.~W.~B.~Muxlow$^{1}$, 
\newauthor
I.~M.~McHardy$^{10}$, D.~R.~A.~Williams$^{10}$, G.~Dumas$^{11}$ \\
$^{1}$ Jodrell Bank Centre for Astrophysics, School of Physics and Astronomy, University of Manchester, Turing Building, \\ Oxford Road, Manchester M13 9PL\\
$^{2}$ College of Astronomy and Space Sciences, University of Chinese Academy of Sciences, Beijing 100049, China\\
$^{3}$ International Centre for Radio Astronomy Research, Curtin University, G.P.O. Box U1987, Perth, WA 6845, Australia\\
$^{4}$ Sydney Institute for Astronomy, School of Physics A28, The University of Sydney, Sydney, NSW 2006, Australia\\
$^{5}$ Jeremiah Horrocks Institute, University of Central Lancashire, Preston PR1 2HE, UK\\
$^{6}$ Cahill Center for Astrophysics, California Institute of Technology, 1216 East California Boulevard, Pasadena, CA 91125, USA\\
$^{7}$ Max-Planck-Institut f\"ur extraterrestrische Physik, Giessenbachstrasse 1, D-85748, Garching bei M\"unchen, Germany\\
$^{8}$ Department of Physics and Astronomy, Hofstra University, New York, USA \\
$^{9}$ Department of Physics and Astronomy, University of Texas-San Antonio, San Antonio, TX 78249, USA\\
$^{10}$ Department of Physics $\&$ Astronomy, University of Southampton, Highfield, Southampton SO17 1BJ, UK\\
$^{11}$ Institut de Radioastronomie Millim\'{e}trique, 300 Rue de la Piscine, F-38406 Saint Martin d'H\'eres, France}

\begin{document}
\setlength{\parskip}{0pt}

\date{Accepted 2018 February 12. Received 2018 February 12; in original form 2017 March 6}

\pagerange{\pageref{firstpage}--\pageref{lastpage}} \pubyear{2014}

\maketitle

\label{firstpage}

\begin{abstract}
We studied the nearby, interacting galaxy NGC\,5195 (M\,51b) in the radio, optical and X-ray bands. We mapped the extended, low-surface-brightness features of its radio-continuum emission; determined the energy content of its complex structure of shock-ionized gas; constrained the current activity level of its supermassive nuclear black hole. In particular, we combined data from the European Very Long Baseline Interferometry Network ($\sim$1-pc scale), from our new e-MERLIN observations ($\sim$10-pc scale), and from the Very Large Array ($\sim$100--1000-pc scale), to obtain a global picture of energy injection in this galaxy. We put an upper limit to the luminosity of the (undetected) flat-spectrum radio core. We find steep-spectrum, extended emission within 10 pc of the nuclear position, consistent with optically-thin synchrotron emission from nuclear star formation or from an outflow powered by an active galactic nucleus (AGN). A linear spur of radio emission juts out of the nuclear source towards the kpc-scale arcs (detected in radio, H$\alpha$ and X-ray bands). From the size, shock velocity, and Balmer line luminosity of the kpc-scale bubble, we estimate that it was inflated by a long-term-average mechanical power $\sim$3--6 $\times 10^{41}$ erg s$^{-1}$ over the last 3--6 Myr. This is an order of magnitude more power than can be provided by the current level of star formation, and by the current accretion power of the supermassive black hole. We argue that a jet-inflated bubble scenario associated
with previous episodes of AGN activity is the most likely explanation for the kpc-scale structures.

\end{abstract}


\section{Introduction}

The early-type galaxy NGC\,5195 (M51b) is the junior partner in the interacting system Messier 51, dominated by the grand-design spiral NGC\,5194 (M\,51a). The peculiar appearance of NGC\,5195 (tidally disturbed in the interaction) has led to discrepant classifications such as SB0$_{1}$-pec 
\citep{SandageTammann1981}, SBa (Hyperleda\footnote{\url{http://leda.univ-lyon1.fr/}}) and I0-pec \citep{RC3}. 

The nature of the nuclear activity of NGC\,5195 has been an outstanding problem. The {\it Spitzer} detection of mid-infrared high excitation lines, such as [Ne {\footnotesize{V}}]$\lambda 14.32 \mu$m, in the inner $\approx(200 \times 400)$ pc, was interpreted \citep{GA2009} as evidence of AGN activity. Optical line ratios also place the nucleus in the LINER class \citep{Moustakasetal10,hoetal1997}, despite the relatively low X-ray luminosity of its nuclear black hole (BH) \citep{Terashima04}.

\begin{figure*}
\centering
\includegraphics[scale=0.317]{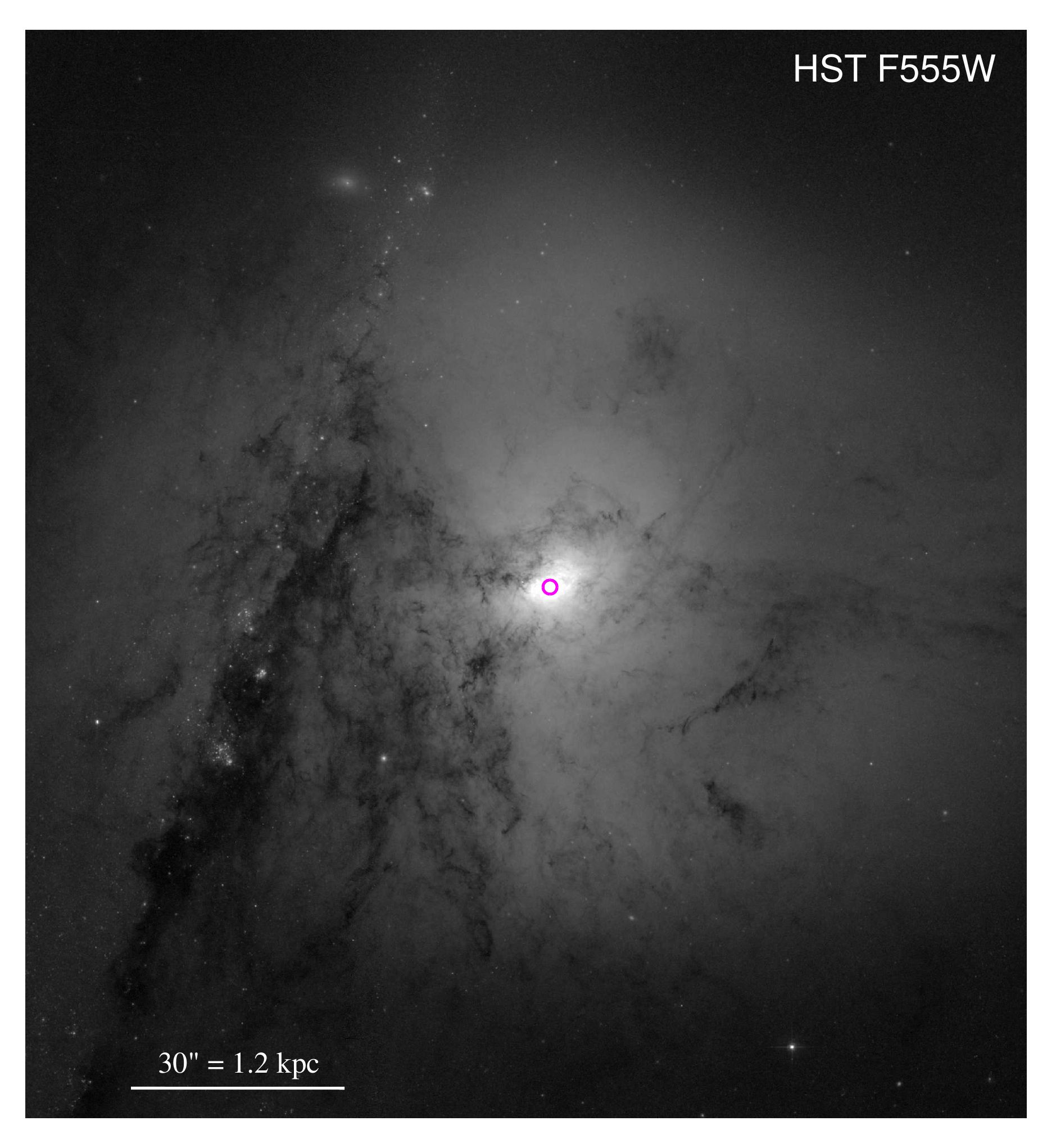}\hspace{-0.35cm}
\includegraphics[scale=0.317]{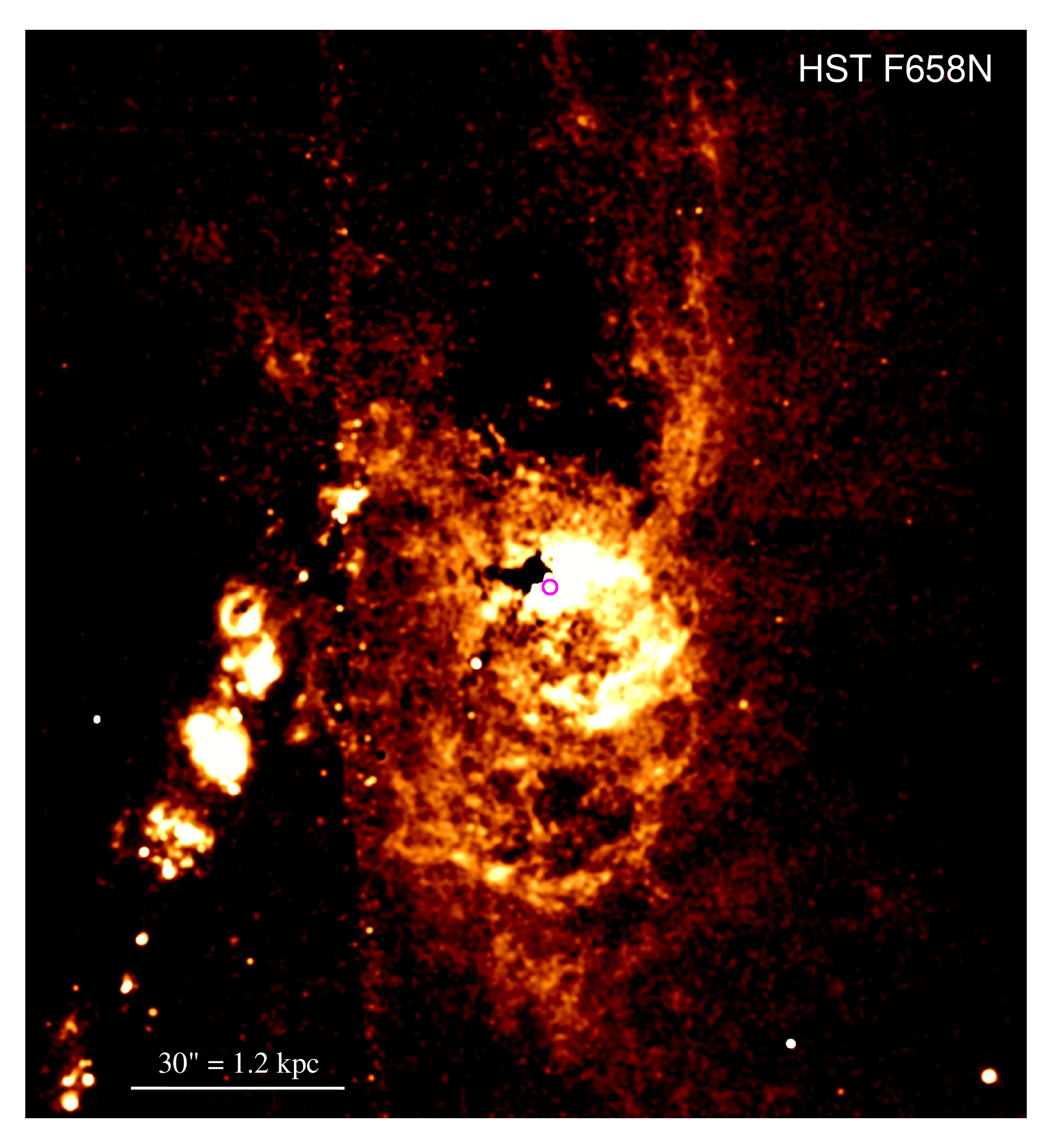}\hspace{-0.35cm}
\includegraphics[scale=0.317]{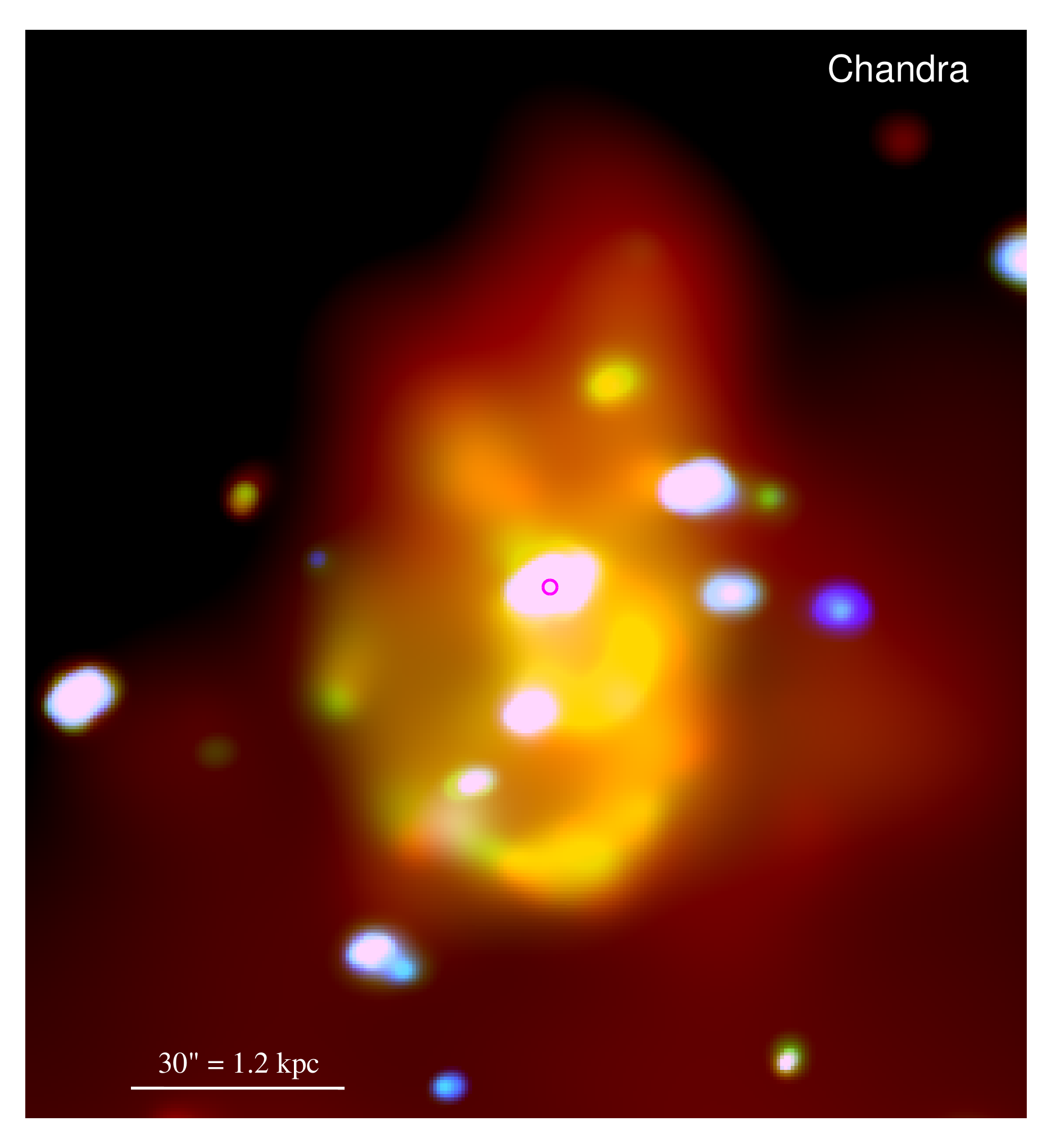}
\caption{Left panel: {\it HST}/ACS image of NGC\,5195 in the F555W filter. The nuclear position is marked with a magenta circle, for comparison with the other two panels. Middle panel: {\it HST}/ACS image in the continuum-subtracted F658N filter (Gaussian-smoothed with an 11-pixel kernel), on the same scale. The bipolar outflow appears open towards the lower-density north side, and bounded by arcs and filaments on the southern side. Right panel: adaptively-smoothed {\it Chandra}/ACIS-S image of the same field. Red $=$ 0.3--1 keV; green $=$ 1--2 keV; blue $=$ 2--7 keV. The morphology of the bipolar bubble is similar to the one seen in the H$\alpha$ image; the faint, diffuse soft emission outside the bubble, in the lower part of the image, belongs to the outer disk of NGC\,5194. In all three panels, north is up and east to the left. The scale bar is 30$''$, corresponding to 1.2 kpc.}   
\label{fig:intro}
\end{figure*}

A recent \textit{Chandra} study by \cite{SJMV} revealed an X-ray emitting region with a double arc-like structure, located on the south side of the galaxy, between about $15''$ and $30''$ from the nucleus.  \cite{SJMV} attributed the X-ray-bright arcs to outgoing shocks, which sweep gas as they radially expand. In addition,  they found that the morphology of the H$\alpha$ emission closely matches that of the X-ray emission (Figure~\ref{fig:intro}), with an H$\alpha$ arc located just outside the outer X-ray arc. \cite{SJMV} suggested that the arcs are the result of episodic outflows driven by mechanical feedback from the nuclear BH; this may have been caused by infalling matter driven by the interaction between NGC\,5195 and NGC\,5194 (interaction timescale of 50--500 Myr: \citealt{SL2000,Dobbsetal2010,Mentuchetal2012}). 


Hydrodynamical simulations show \citep{Saxton+2005, Massaglia+2016,Mukherjee+2017} that the lack of obvious jet signatures such as hot spots is expected for a system with low jet power ($P_{\rm jet} \la 10^{42}$ erg s$^{-1}$) and/or a jet propagating in an inhomogeneous medium. In those cases, the jet entrains ambient gas and gradually dissipates its kinetic energy along the way, losing collimation and forming a bubble or cocoon, with an arc-like front. Weak jets are also subject to the magnetic kink instability \citep{Tchekhovskoy2016}, which favours the dissipation of the jet power in a bubble. (See also the review of jets and bubbles by \citealt{Soker2016}.) On the other hand, starburst-driven superwinds can also produce large galactic bubbles and bipolar outflows ({\it e.g.}, \citealt{Baum+1993, Jogee+1998, Hota+Saikia2006}).

In this paper, we want to determine whether the arc structures in NGC\,5195 are more likely to have a starburst or a jet origin. To do so, we follow up on the findings of \cite{SJMV}, by investigating the connection between the current nuclear activity and the kinetic energy required to create the arcs/bubble structure. We present the results of our new observations with the enhanced Multi-Element Remote-Linked Interferometer Network (e-MERLIN) at 18cm and 6cm, as part of the Legacy 
e-MERLIN Multi-Band Imaging of Nearby Galaxies Survey (LeMMINGS) project \citep{baldi2018,lemmings2014}. We combine those observations with our data from the European VLBI (very long baseline interferometry) Network (EVN), and with archival Very Large Array (VLA) observations, in order to map the radio properties from the nuclear region (pc scale) to the arc region (kpc scale), and determine how energy was injected from the former to the latter region. We support our analysis with a {\it Chandra} study of the X-ray emission from the nuclear region, and with {\it Hubble Space Telescope} ({\it HST}) narrow-band imaging of the Balmer line emission in the arc region.

Throughout this paper we define spectral index, $\alpha$, as $S_{\nu} \propto \nu^{\alpha}$, where $S_{\nu}$ is the measured flux density at frequency, $\nu$. For consistency with \cite{SJMV}, we adopt a distance to NGC\,5195 of 8.0 Mpc (thus, $1' \approx 40$ pc).

\section{Observations and Data Analysis}
\label{sec:obs}

\subsection{European VLBI Network data}
We observed NGC\,5195 (RA $= 13^{\mathrm{h}}$29$^{\mathrm{m}}$59.537$^{\mathrm{s}}$; Dec $= +47^{\circ}$15$^{\prime}$58$^{\prime\prime}$.38 [J2000.0]) with the EVN on 2011 November 7, for 8 hr (UTC: 01:30--09:30) at a wavelength of 18cm, with a single target pointing centred on NGC\,5194. Details of the observing set-up are given in \citet{Rampadarathetal15}.

The observation was phase referenced to J1332$+$4722 (RA = 13$^{\mathrm{h}}$ 32$^{\mathrm{m}}$ 45.246424$^{\mathrm{s}}$; Dec = +47$^{\circ}$ 22$^{\prime}$ 22$^{\prime\prime}$.667700 [J2000.0]) located 31$^{\prime}$ from NGC\,5194, and used a phase-referencing cycle of 5 minutes on target and 1 minute on the calibrator.  Using the multiple-phase-correlation technique \citep{Delleretal11,Morganetal11}, a field of $\approx$ 8\arcm was surveyed using 192 sub-fields, including one centred on NGC\,5194. Each sub-field covered 41\arcs $\times$ 41\arcs, at a maximum resolution of 5 mas. The phase centres, including NGC\,5194 were calibrated and primary beam corrected using the Astronomical Image Processing System (\textsc{aips})\footnote{The Astronomical Imaging Processing System is owned and maintained by the National Radio Astronomical Observatory (NRAO)} as described by \citet{Rampadarathetal15}. Images of the NGC\,5194 centred field were made with different combinations of UV-ranges (10$\%$ to 100$\%$ ) and image weights, which did not reveal any detections.  Using the most sensitive image, we place a 5-$\sigma$ upper limit on the flux density of $F_{\nu} < 132 \mu$Jy beam$^{-1}$.

\subsection{e-MERLIN data}

NGC\,5195 was observed with e-MERLIN as part of the LeMMINGS project, at both L and C bands\footnote{The L-band covers 1--2 GHz; the C-band 4--8 GHz; the X-band 8--12 GHz.}. The observations are summarised in \Tab~\ref{tab:eMerObs}. 
The L-band observation was phase referenced to J1335$+$4542 (RA $= 13^{\mathrm{h}}$35$^{\mathrm{m}}$21$^{\mathrm{s}}$.9623; Dec $= +45^{\circ}$42$^{\prime}$38$^{\prime\prime}$.231 [J2000.0]), located 1$^{\circ}$.81 from our target.
The C-band observation was phase-referenced to the nearby calibration source J1332$+$4722 (RA $= 13^{\mathrm{h}}$32$^{\mathrm{m}}$45$^{\mathrm{s}}$.246; Dec $= +47^{\circ}$22$^{\prime}$22$^{\prime\prime}$.667 [J2000.0]) located 0$^{\circ}$.48 from NGC\,5195. All observations used a phase-referencing cycle of 7 minutes on target and 1 minute on the calibrator. We observed the standard flux calibrator 3C\,286 to calibrate the flux density scale to an accuracy of 5$\%$, and used J1407$+$2827 (OQ\,208) to calibrate the bandpass.

Both e-MERLIN datasets were obtained with 512 channels per intermediate frequency (four in the C-band, eight in the L-band), and were averaged to 128 channels per intermediate frequency in post-processing. Following extensive flagging of bad data due to radio frequency interference and telescope dropouts, we calibrated the datasets with 
\textsc{aips}, using the e-MERLIN calibration pipeline\footnote{\url{http://www.e-
merlin.ac.uk/observe/pipeline/}} \citep{Argo2014}, with the procedure described in the e-MERLIN 
Cookbook\footnote{\url{http://www.e-merlin.ac.uk/data_red/tools/e-merlin-cookbook_V3.0_Feb2015.pdf}}. 

\begin{figure*}
\centering
\includegraphics[scale=0.5,clip=true,trim=0cm 0.1cm 0cm 0cm]{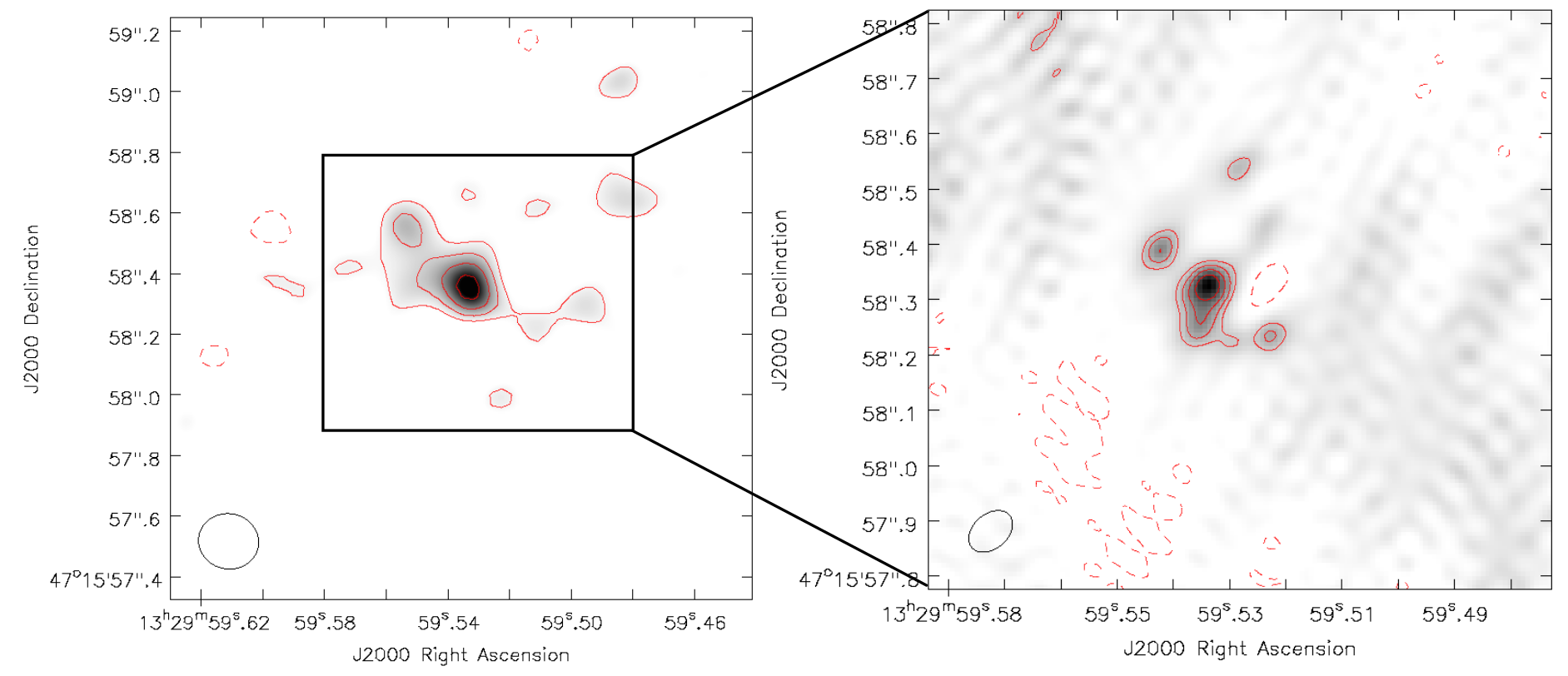}
\caption{e-MERLIN maps of the nuclear region of NGC\,5195. Left panel: L-band contours overlaid on the corresponding greyscale image; the image size is $1''.6 \times 1''.8$ (64 pc $\times$ 72 pc). The synthesized beam is $0''.20 \times 0''.18$ (8.0 pc $\times$ 7.2 pc) at a position angle of $79^{\circ}$. Right panel: C-band contours and greyscale image; the image size is $1''.0 \times 1''.0$ (40 pc $\times$ 40 pc). The synthesized beam is $0''.09 \times 0''.06$ (3.6 pc $\times$ 2.4 pc) at a position angle of $-49^{\circ}$. In both panels, the contours are at $(-3, 3, 4, 5, 6, 7)  \times \sigma_{\rm{rms}}$ (as listed in \tab~\ref{tab:fluxMeas}); negative contours are shown as  dashed lines.}   
\label{fig:MERLINimg}
\end{figure*}

To make full use of the wide-bandwidth e-MERLIN data, we imaged the calibrated datasets with the multi-scale, multi-frequency synthesis algorithm \citep{RC2011} in the Common Astronomical Software 
Applications (\textsc{casa}; \citealt{McMullin+07}). To search for emission from the VLA-detected structures (see Section~\ref{sec:vla_cont}), we used standard wide-field imaging techniques, accounting for non-coplanar baselines (w-projection with 64 w-planes{; see} \citealt{Cornwelletal2008}). We did not apply any time or spectral averaging, in order to minimise amplitude losses due to time and bandwidth smearing. Only emission from the core region was detected above 5$\sigma$. The final  naturally-weighted images of the core region at both frequencies are displayed in \fig~\ref{fig:MERLINimg}. While no self-calibration was used on the target, improved solutions for the images were obtained via phase-only self-calibration of the phase calibrator and interpolated to NGC\,5195 data. 

We measured flux densities using \textsc{blobcat} \citep{Hales2012_paper,Haless12_soft}, a stand-alone Python program to catalogue objects in a 2D-radio-astronomy Stokes-I image. \textsc{blobcat} presents a robust alternative to Gaussian fitting by utilizing the flood fill algorithm \citep{magnenat2012computer} to detect and catalogue complex ``blobs" that represent sources in a 2D-astronomical image. To prevent overfitting of the blobs, we used a stopping signal-to-noise ratio\footnote{The parameter $f_{\rm{SNR}}$ in \textsc{blobcat}.} of 3.


\begin{table}\footnotesize
\centering
\caption{e-MERLIN Observations}
\label{tab:eMerObs}
\begin{tabular*}{0.48\textwidth}{l @{\extracolsep{\fill}} cccc}
\hline\hline
\noalign{\smallskip}
 Band & $\nu$ & $\Delta\nu$ & Date & Time  \\ 
 & (GHz) & (MHz) & (dd/mm/yy)  & (hr) \\ 
(1)   & (2)   &  (3) & (4) & (5)\\[5pt] 
 \hline \\[-5pt]
L & 1.51 & 512 & 09/04/15 & 0.90 \\
C & 5.07 & 512 & 08/03/16 & 5.85 \\[5pt]
\hline
\hline
\end{tabular*}
\begin{flushleft}
{\bf {Notes.}} Columns are: (1) band name; (2) central frequency; (3) bandwidth; (4) observing dates; (5) total on-source integration time.
\end{flushleft}
\end{table}

\subsection{Archival VLA data} 
\label{sec:vla_cont}
\begin{figure*}\scriptsize
\centering
\subfloat[L~Band]{\includegraphics[scale=0.43,clip=true,trim=0cm 2cm 1cm 0cm]{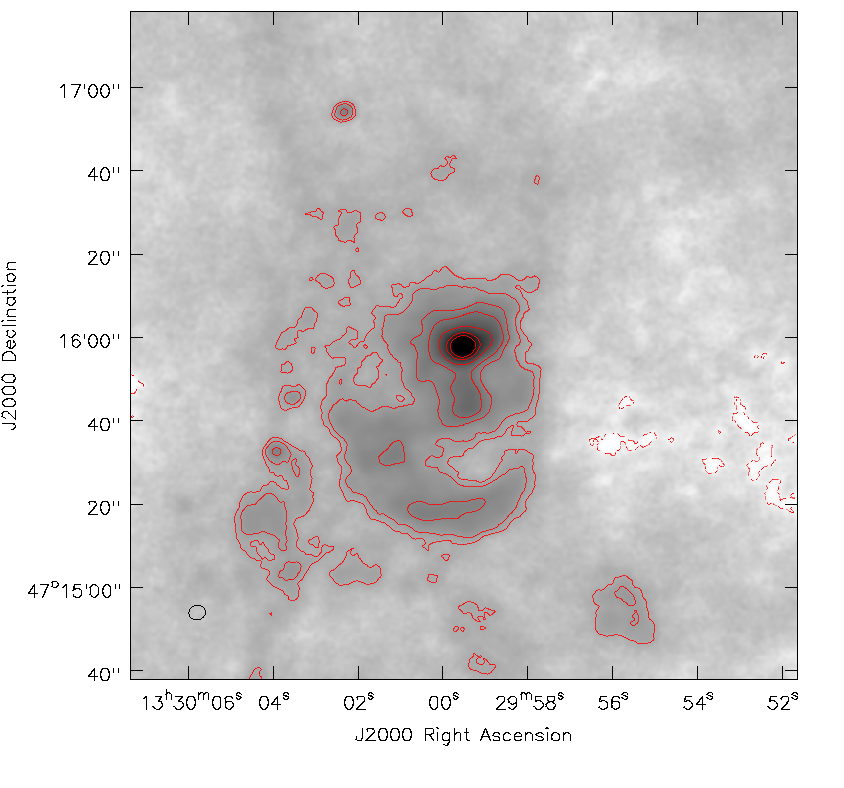}}
\subfloat[C~Band]{\includegraphics[scale=0.43,clip=true,trim=0cm 2cm 1cm 0cm]{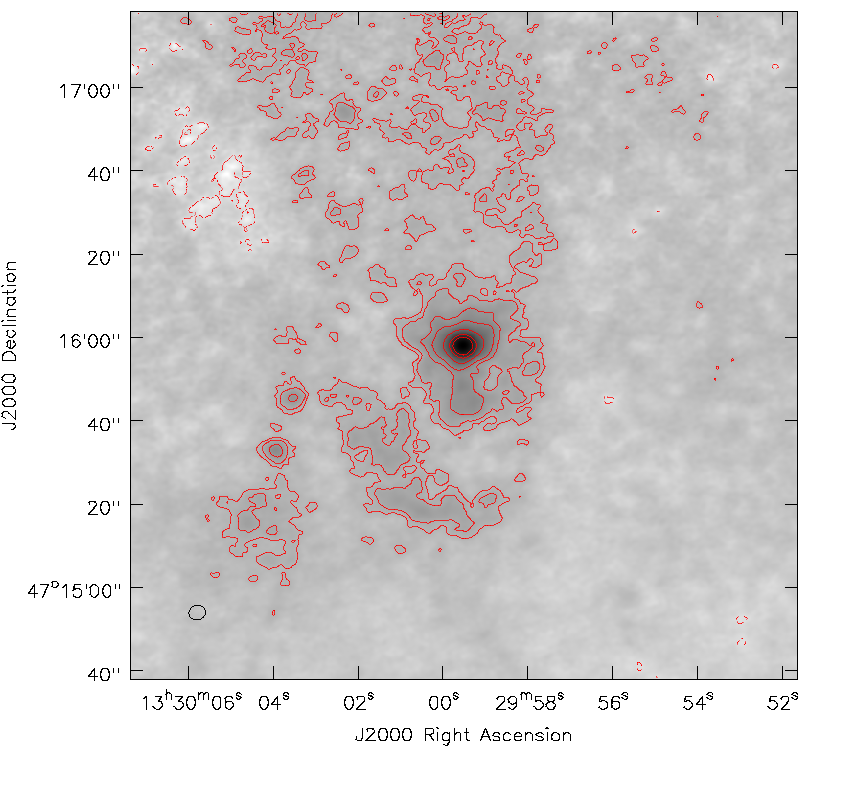}}\\
\subfloat[X~Band]{\includegraphics[scale=0.43,clip=true,trim=0cm 2cm 1cm 0cm]{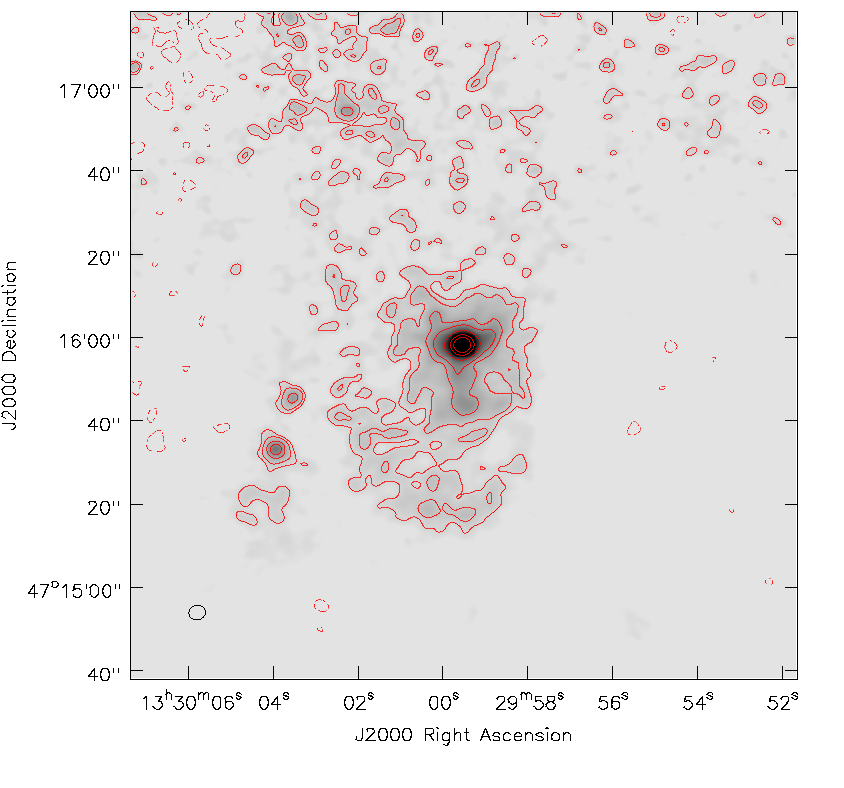}}
\caption{Multi-scale, multi-configuration, multi-frequency VLA images of NGC\,5195. The peak fluxes and the image noise, $\sigma_{rms}$, are given in \tab~\ref{tab:fluxMeas}. All three images have a size of $2'.5 \times 2'.5$ (6~kpc $\times$ 6~kpc), with synthesized beam sizes of $4''.01 \times 3''.34$ (160 pc $\times$ 138 pc), at a position angle of $-79^{\circ}$. The contours in each image are at $(-3, 3, 5, 10, 15, 30, 60, 90) \times \sigma_{\rm{rms}}$; negative contours are dashed. All three images show a core, a jet-like elongated structure south of the core (``spur"), and an arc-like feature further south.}
\label{fig:VLAMaps}
\end{figure*}

We used archival multi-configuration VLA observations centred on NGC\,5194 in the L, C and X bands.  The observations are summarised in \Tab~\ref{tab:vlaObs}, and the calibration procedures are explained in detail by \cite{Dumasetal2011}. { The uv datasets for each band were combined within \textsc{casa} 
and imaged using multi-scale clean \citep{RC2011}, with scales at 0, 5, 10 and 15 pixels. }For 
accurate morphological and flux density comparisons, the final images (at each frequency) were made 
with minimum and maximum uv-ranges of 0.5~k$\lambda$ and 180~k$\lambda$ respectively, a cellsize of 
$0''.23$, with a Briggs robust weighting of 1. Since the datasets were all centred on NGC\,5194, in order to include NGC\,5195 ($\approx$5$'$ north of the pointing centre), we created wide-field images of radius $6'$, using w-projection \citep{wproj} with 128 planes. Primary beam effects were corrected with the \textsc{casa} task \textsc{impbcor}, and finally a sub-image of NGC\,5195 was extracted with \textsc{imsubimage}. The final result is shown in \fig~\ref{fig:VLAMaps}. As for the e-MERLIN data, we used \textsc{blobcat} to extract flux densities. For further analysis, we also used other Python-based open-source software such as Astropy\footnote{\url{http://www.astropy.org/}} 2.0, and SciPy\footnote{\url{https://scipy.org/}} 1.0 (in particular, its core package NumPy).

\begin{table}
\centering
\caption{Archival VLA Observations}
\label{tab:vlaObs}
\begin{tabular*}{0.48\textwidth}{l @{\extracolsep{\fill}} cccc}
\hline\hline\\[-5pt]
Band & Config & $\nu$ &    Date     & Time \\
     &               & (GHz) &   & (hr) \\
(1)  &      (2)      & (3)   &   (4)       &  (5)   \\[5pt]
 \hline\hline\\[-5pt]
L & A & 1.385$^{*}$;1.465$^{*}$ & 10/2004 & 30.8 \\
  & B & 1.385$^{*}$;1.465$^{*}$ & 06/2005   & 23.9 \\
  & C & 1.465$^{\dagger}$ & 04/1992  & 13.4 \\
  & D & 1.465$^{\dagger}$ & 07/1988 & 12 \\[5pt]
\hline\\[-5pt]
C & B & 4.876$^{*}$ & 12/2003-03/2005 & 23 \\
  & C & 4.835$^{\dagger}$ & 07/2001 & 14.4 \\
  & D & 4.885$^{\dagger}$ & 03-04/1991 & 8.7 \\[5pt]
\hline\\[-5pt]
X & C & 8.435$^{\dagger}$ & 04/2005 & 20.3 \\
  & D & 8.435$^{\dagger}$ & 10-11/2001 & 17.6 \\[5pt]
\hline\hline
\end{tabular*}
    \begin{flushleft}
    {\bf{Notes.}} See \citealt{Dumasetal2011}, Table 2, for more details. \\
    Columns are: (1) band name; (2) VLA configuration; (3) central observing frequency for each intermediate frequency; (4) observation dates (mm/yyyy); (5) total on-source integration time.\\
    $^{\dagger}$ bandwidth = 50 MHz; $^{*}$ bandwidth = 25 MHz.
    \end{flushleft}
\end{table}

\begin{table*}
\centering
\caption{Radio Interferometric Measurements}
\label{tab:fluxMeas}
\begin{tabular*}{\textwidth}{l @{\extracolsep{\fill}} cccccccc}
\hline\hline\\[-5pt]
Array & $\nu$ & Component & $S_{\rm{peak}}$ & $S_{\rm{int}}$ & log($L_{\nu}$/W~Hz$^{-1}$) & $\sigma_{\rm{rms}}$ & $\Theta_{\rm{beam}}$  & P.A.\\ 
      & (GHz) & & (\mjybm) & (mJy)	 &  & (\mujybm) & (\arcs $\times$ \arcs)  & ($^{\circ}$)\\ 
(1)   & (2)   & (3) 	& (4) 		 & (5)  & (6) & (7) & (8) & (9)\\ [5pt]
\hline\hline\\[-5pt]
EVN  & 1.65    &   core  &   $<$0.132    &  $<$0.132      &    $<$18.00      &    26.4     &   0.0099 $\times$ 0.0056     &     14.3      \\[5pt]
 \hline \\[-5pt]
e-MERLIN & 1.51 & core & 0.57 $\pm$ 0.11 & 2.00 $\pm$ 0.17 & 19.18 & 98.90 & 0.20 $\times$ 0.18 & 79 \\
         \cline{2-9}\noalign{\smallskip} 
 & 5.07 & core & 0.18 $\pm$ 0.03 & 0.42 $\pm$ 0.04 & 18.51 & 26.90 & 0.09 $\times$ 0.06 & -49 \\[5pt]
\hline\\[-5pt]
VLA & 1.43 & total & 5.31 $\pm$ 0.39 & 39.03 $\pm$ 2.76 & 20.48 & 24.42 & 4.01 $\times$ 3.45 & -79\\
         & &  core & 5.31 $\pm$ 0.39 & 11.22 $\pm$ 0.79 & 19.93 & $\cdots$ & $\cdots$ & $\cdots$ \\
         & &  ``spur" & $\cdots$ & 7.10 $\pm$ 0.85  & 16.74 & $\cdots$ & $\cdots$ & $\cdots$ \\
         & &  arc & $\cdots$ & 18.14 $\pm$ 2.18 & 17.14 & $\cdots$ & $\cdots$ & $\cdots$ \\
         \cline{2-9}\noalign{\smallskip} 
         
 & 4.86 & total & 2.44 $\pm$ 0.18 & 10.25 $\pm$ 0.73 & 19.89 & 14.13 & 4.01 $\times$ 3.45 & -79\\
 &  & core & 2.44 $\pm$ 0.18 & 4.26 $\pm$ 0.30 & 19.51  & $\cdots$ & $\cdots$ & $\cdots$\\ 
 &  & ``spur" & $\cdots$ & 2.97 $\pm$ 0.35 & 16.36  & $\cdots$ & $\cdots$ & $\cdots$\\
 &  & arc & $\cdots$ & 3.02 $\pm$ 0.36 & 16.36 & $\cdots$ & $\cdots$ & $\cdots$\\
 \cline{2-9}\noalign{\smallskip} 
 
 & 8.44 & total & 1.78 $\pm$ 0.13 & 9.51 $\pm$ 0.67 & 19.86 & 11.56 & 4.01 $\times$ 3.45 & -79 \\
 & & core & 1.78 $\pm$ 0.13 & 2.68 $\pm$ 0.19 & 19.31 & $\cdots$ & $\cdots$ &$\cdots$ \\
 & & ``spur" & $\cdots$ & 2.51 $\pm$ 0.30 & 16.28 & $\cdots$ & $\cdots$ &$\cdots$ \\
 & & arc & $\cdots$ & 4.32 $\pm$ 0.30 & 16.52 & $\cdots$ & $\cdots$ &$\cdots$ \\[5pt]
\hline
\hline
\end{tabular*}
    \begin{flushleft}
    {\bf {Notes.}} Columns are: (1) array; (2) central frequency; (3) spatial components, as defined in \sect~\ref{sec:arcsec_scales}; (4) peak flux density; (5) integrated flux density; (6) spectral luminosity corresponding to the integrated flux density at 8 Mpc; (7) root-mean-square noise; (8) beam size; (9) beam position angle.
    \end{flushleft}
\end{table*}

\subsection{Optical data}

We used optical images of the galaxy taken with the {\it HST} Advanced Camera for Surveys (ACS) on 2005 January 12--22 (observing program 10452); we downloaded the mosaiced images from the Hubble Legacy Archive\footnote{\url{https://hla.stsci.edu/}}.
We used a suitably weighted linear combination of the F555W and F814W broad-band images to construct an approximate continuum image at the wavelength of H$\alpha$; we then subtracted it from the F658N narrow-band filter image, after a suitable re-normalization based on their effective filter widths. We verified that the main continuum sources (stars and clusters) were, on average, well subtracted; thus, the continuum-subtracted F658N image (Figure~\ref{fig:intro}, middle panel) includes only the H$\alpha$ plus [N {\footnotesize II}] emission lines complex.

\subsection{X-ray data}
In the X-ray band, the M\,51 system has been observed 12 times (considering only observations longer than 2 ks) between 2000 and 2012, with the back-illuminated S3 chip of the Advanced CCD Imaging Spectrometer (ACIS-S) on board the \textit{Chandra X-ray Observatory} (see Table 3 in \citealt{Rampadarathetal15}). The total ACIS-S exposure time for the central region of NGC\,5195 was 658 ks.
It was then observed once with the front-illuminated chips of \textit{Chandra}/ACIS-I, on 2017 March 17, for 38 ks; the Principal Investigator (PI) was Murray Brightman. 
The ACIS-S datasets were downloaded from the public {\it Chandra} archive. We reprocessed all observations using tools available in {\small {CIAO}} version 4.9, with the {\small{CALDB}} calibration files version 4.7.7 \citep{CAIO_paper}, to obtain new level-2 event files. We reprojected the event files to a common point, and combined them, to obtain a deep X-ray image of the field, as described in detail by \cite{Rampadarathetal15}. 
The ACIS-I dataset will be discussed in more details in a forthcoming paper (Brightman et al., in preparation; see also \citealt{Brightman2017}).

X-ray emission from the nuclear region of NGC\,5195 is detected in every ACIS-S and ACIS-I observation.
Unfortunately, in the ACIS-S observations, NGC\,5195 was always located several arcmin away from the S3 aimpoint (typically, the nuclear region of NGC\,5194); as a result, the point-spread-function near the nucleus is highly elongated (Figure~\ref{fig:intro}, right panel) and the spatial resolution relatively low, $\approx$5$''$. Also, for this reason, we cannot improve on the default {\it Chandra}/ACIS-S astrometry of the re-processed and stacked archival data; the astrometry is accurate within $\approx$0$^{\prime\prime}$.5 in 68\% of the imaging datasets, and within $\approx$0$^{\prime\prime}$.7 in 90\% of the datasets\footnote{\url{http://cxc.harvard.edu/cal/ASPECT/celmon/}}.
Instead, the 2017 ACIS-I observation was aimed at a source in the northern half of NGC\,5194: hence, the resolution around NGC\,5195 is better, $\approx$1$''$. This is enough to (partially) resolve the nuclear emission into distinct sources; however, the exposure time is too short for any meaningful spectral analysis. For ACIS-I observations, the 90\% uncertainty of the absolute position has a radius of $\approx$0$^{\prime\prime}$.75. We checked the astrometric accuracy of our observation with the help of four radio/X-ray associations, selected from the list of sources in Table 4 of \cite{Rampadarathetal15}. There is no systematic offset between the X-ray and radio positions, and the scatter is $\la$0$''$.5.

We used a combination of the deep (but low-resolution) ACIS-S archival data and the shallow (but high-resolution) ACIS-I data to estimate the X-ray properties of the nucleus.
For the ACIS-S data, we used the imaging tool {\small{ds9}} to define an elliptical source region of semi-major axes $4''.1 \times 2''.8$ ($\approx$160 pc $\times$ 110 pc), and a background region three times as large. We then used standard {\small CIAO} tasks for data analysis. In particular, 
we measured the count rate and hardness ratio of the unresolved nuclear source at every epoch. Then, we extracted a merged spectrum of all the observations, with associated response and auxiliary response files, using the {\small CIAO} task {\it specextract}. We grouped the spectrum to a minimum of 15 counts per bin, with the {\it grppha} task within {\small FTOOLS} \citep{1995ASPC...77..367B}.  Finally, we modelled the spectrum with {\small XSPEC} Version 12.9.0.
For the ACIS-I observation, we identified a candidate point-like nuclear source, consistent with the location of the radio nucleus, and estimated its net count rate. We converted this count rate to an unabsorbed flux (and hence luminosity) in various bands, using plausible spectral parameters inferred from the analysis of the stacked ACIS-S spectrum (as discussed in Section 3.4).

\begin{figure*}
\centering
\hspace{-0.15cm}
\includegraphics[scale=0.475]{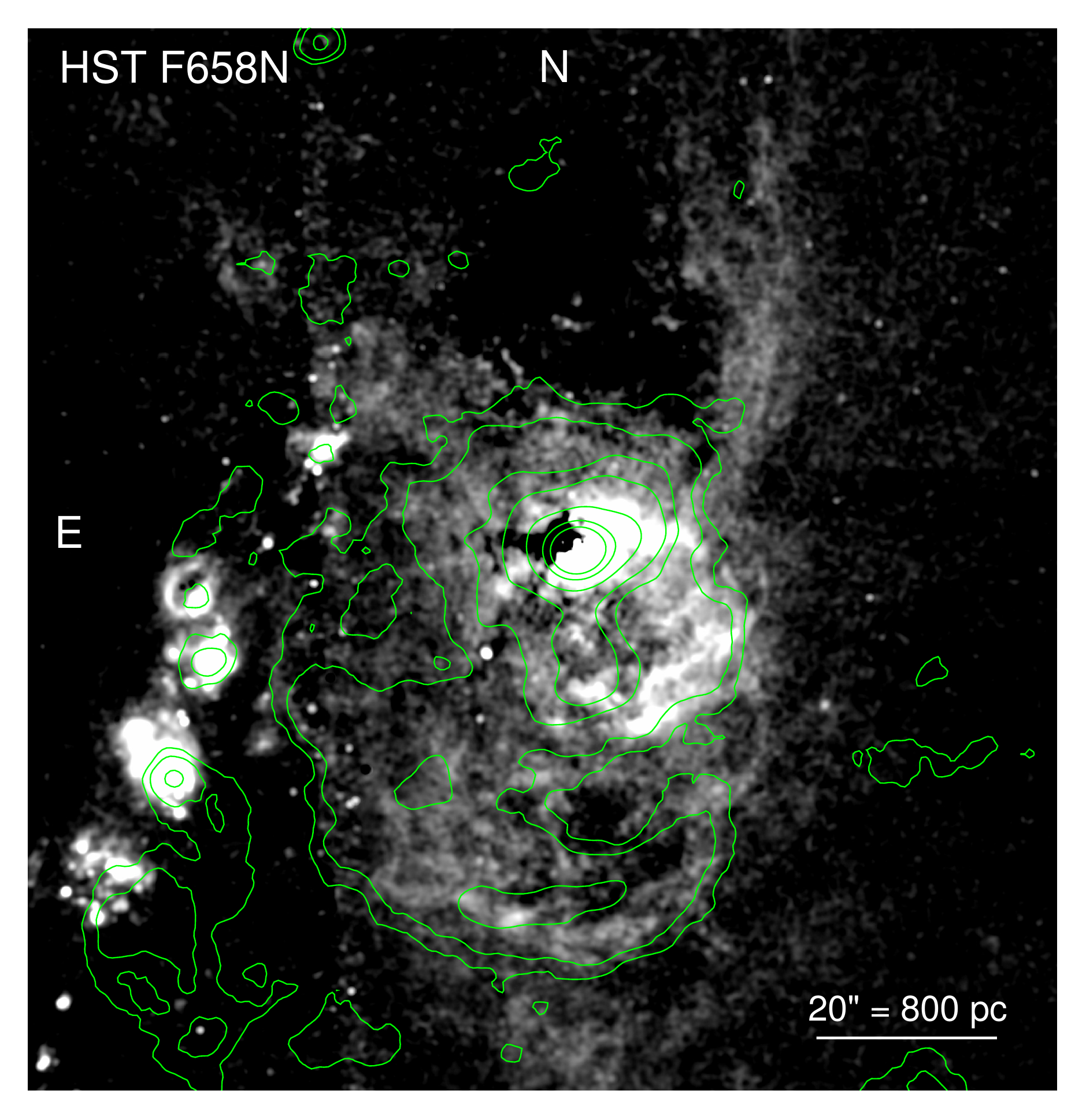}\hspace{-0.5cm}
\includegraphics[scale=0.475]{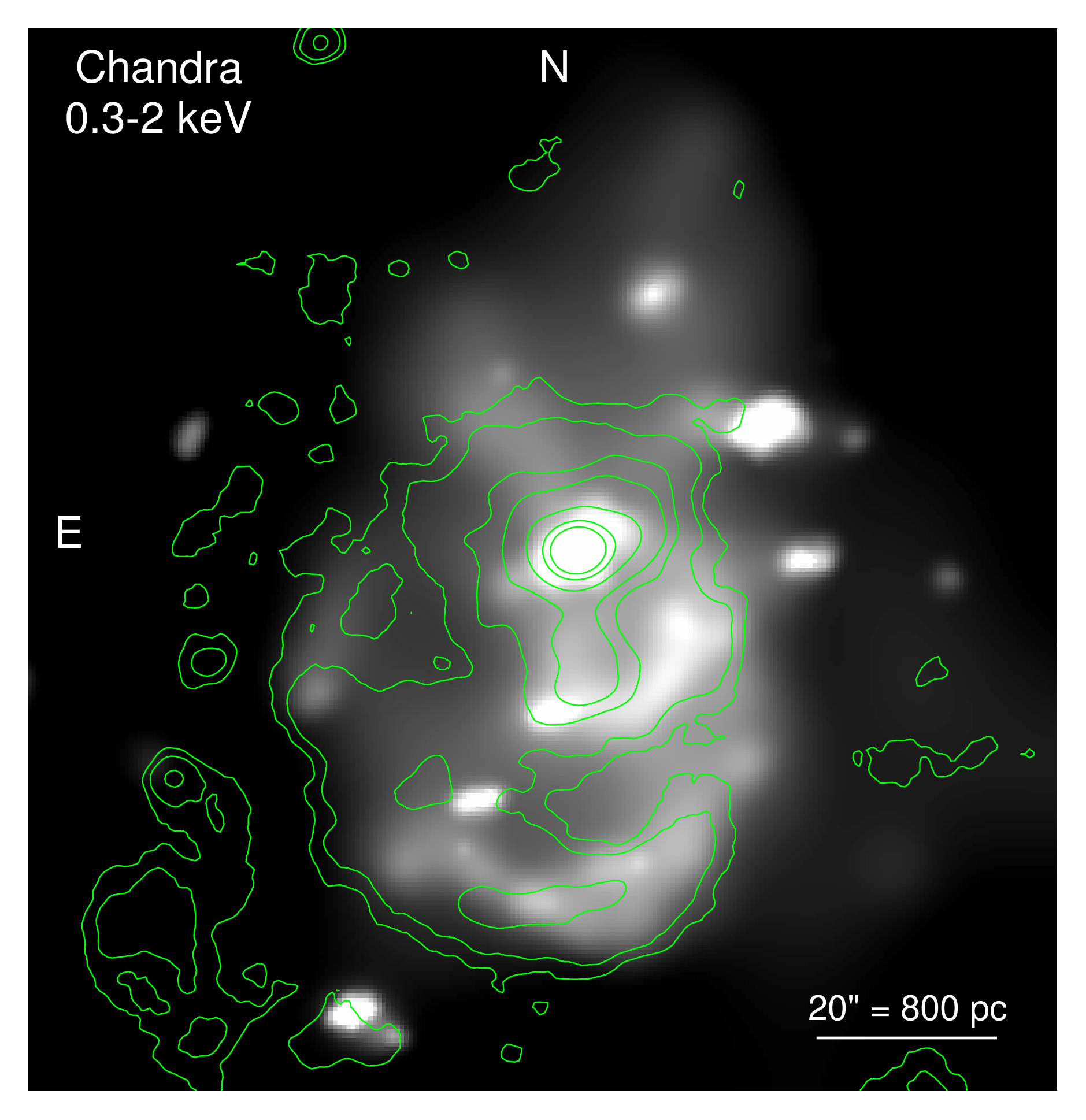}
\vspace{-0.4cm}
\caption{Left panel: VLA 20-cm contours (in green) overplotted onto a greyscale {\it HST}/ACS image in the continuum-subtracted F658N band. The {\it HST} image has been Gaussian smoothed with an 11-pixel kernel. The radio contours correspond to those plotted and explained in Figure 3a. The scale bar is 20$''$, corresponding to $\approx$800 pc, and the size of the panel is $\approx$1.9$'$ $\times$ 2$'$.
Right panel: VLA 20-cm contours (in green) overplotted onto a greyscale {\it Chandra}/ACIS-S image in the 0.3--2 keV band. The {\it Chandra} image has been adaptively smoothed with the CIAO task {\it{csmooth}}. We selected the 0.3--2 keV band in order to maximize the relative contribution of the diffuse hot gas emission. Scale bar and image size are the same as in the left panel.}   
\label{fig4new}
\end{figure*}

\section{Results}

\subsection{Sub-arsecond-scale radio structure}

We do not detect any emission at the nuclear position in the EVN images, at 18cm; the 5-$\sigma$ upper limit of the flux density is $F_{\nu} < 132$ $\mu$Jy beam$^{-1}$ \citep{Rampadarathetal15}, with a resolution of 9.9 mas $\times 5.6$ mas (corresponding to 0.38 pc $\times$ 0.22 pc). This implies a 5-GHz upper limit $L_{\rm{5GHz}} < 5.1 \times 10^{34}$ erg s$^{-1}$ for the luminosity of a compact, flat-spectrum AGN jet in November 2011. It is possible that we did not detect a compact source with the EVN because the jet is currently switched off, or because the source is dominated by a larger-scale radio outflow.

Instead, we do detect nuclear radio emission in the e-MERLIN observations of 2014--2015, in both the L and C bands. Initially, we imaged the e-MERLIN data with natural weighting; the beam sizes are 200 mas $\times 180$ mas at 1.5 GHz, and 90 mas $\times 60$ mas at 5 GHz. The images show (Figure \ref{fig:MERLINimg}) a source with a stronger core and a resolved structure on scales of about $0''.3$ (12 pc) in the L band\footnote{The source extents are taken as the linear distance from the central peak pixel to the edge of the 3-$\sigma_{rms}$ contour. A single pixel accuracy is assumed for this measurement.}. The core is itself partially resolved in the C band, with a size of about $0''.1$ (4 pc). The nuclear emission has a steep spectrum, with integrated flux densities of $(2.00 \pm 0.17)$ mJy at 1.5 GHz and $(0.42 \pm 0.04)$ mJy at 5 GHz (Table \ref{tab:fluxMeas}). The uncertainties are a combination of the fitting errors from {\it blobcat} that take into account correlated pixel noise, and of a systematic flux calibration error of $5\%$, added in quadrature.

We then re-imaged the e-MERLIN C-band data with a Briggs robust weighting of $-1$, which provides a higher spatial resolution:
40.8 mas $\times 28.5$ mas (1.6 pc $\times 1.1$ pc). At those scales, we find that the nuclear source is unresolved. 
The location of the nuclear source is RA $= 13^{\mathrm{h}}$29$^{\mathrm{m}}$59$^{\mathrm{s}}$.534 ($\pm 0^{\mathrm{s}}$.001); Dec $= +47^{\circ}$15$^{\prime}$58$^{\prime\prime}$.33 ($\pm 0^{\prime\prime}$.01) [J2000.0].

We used the \textsc{aips} task {\it jmfit} to measure the flux density and an upper limit on the deconvolved size. We obtain a 5-GHz flux density of $(0.28 \pm 0.07)$ mJy and an upper limit on the deconvolved size of $0''.033 \approx 1.3$pc. This implies a brightness temperature $\log(T_{\rm B}/{\rm K}) \ga 4.1$. Generally, thermal sources have $\log(T_{\rm B}/{\rm K}) \la 5.0$ and  non-thermal sources have $\log(T_{\rm B}/{\rm K}) \ga 5.0$ \citep{Condon1992}. Therefore, the limit on the brightness temperature in the core of NGC\,5195 is low enough to be still consistent with the starburst interpretation (although of course it does not rule out an AGN interpretation).
Note that this 0.28-mJy detection is not inconsistent with the 0.13-mJy upper limit from the EVN, because the beam area is 20 times smaller in the latter case.
Deep radio continuum observations at higher angular resolution would be needed to further constrain whether the nuclear radio source is thermal or non-thermal in nature.





\subsection{Arcsecond-scale radio structure}
\label{sec:arcsec_scales}

The VLA images (Figure \ref{fig:VLAMaps}) reveal a number of interesting features: (1) a bright compact source at the centre of the galaxy; (2) an elongated structure that extends about $18''.5$ (0.74 kpc) south of the nucleus (hereafter referred to as the ``radio spur"); and (3) an arc of emission at about $46''.3$ (1.85 kpc) south of the nucleus that appears to curve around the nuclear position (hereafter referred to as the ``radio arc"). Flux densities and corresponding uncertainties for the VLA images are listed in \Tab~\ref{tab:fluxMeas}. 

There is also evidence for diffuse, kpc-scale low-surface brightness emission to the north of the nuclear location, that appears to create a large elongated cocoon, completely surrounding the central region of the galaxy; see also the top two panels of Fig.~2 in \citet{Dumasetal2011}, where the cocoon appears to extend at least $\approx$3 kpc in the northern direction, twice as far as the southern lobe. It is difficult to say with confidence whether this is real emission, and whether it has a well-defined boundary, as the structure is not consistent across all frequencies. In addition, the distance of NGC\,5195 from the pointing centre of the VLA observations (i.e. at the outer edges of the primary beam) increases the image rms noise and hence the uncertainty of the emission being real. With those caveats, the emission in the northern direction is only a 3-$\sigma$ detection, but it is an intriguing possibility that deserves further investigations. The presence of an elongated radio cocoon around the nucleus would be a hint that kinetic power has been injected from the nucleus into the surrounding gas in a bipolar geometry.

\begin{figure}
\centering
\hspace{-0.1cm}
\includegraphics[scale=0.315]{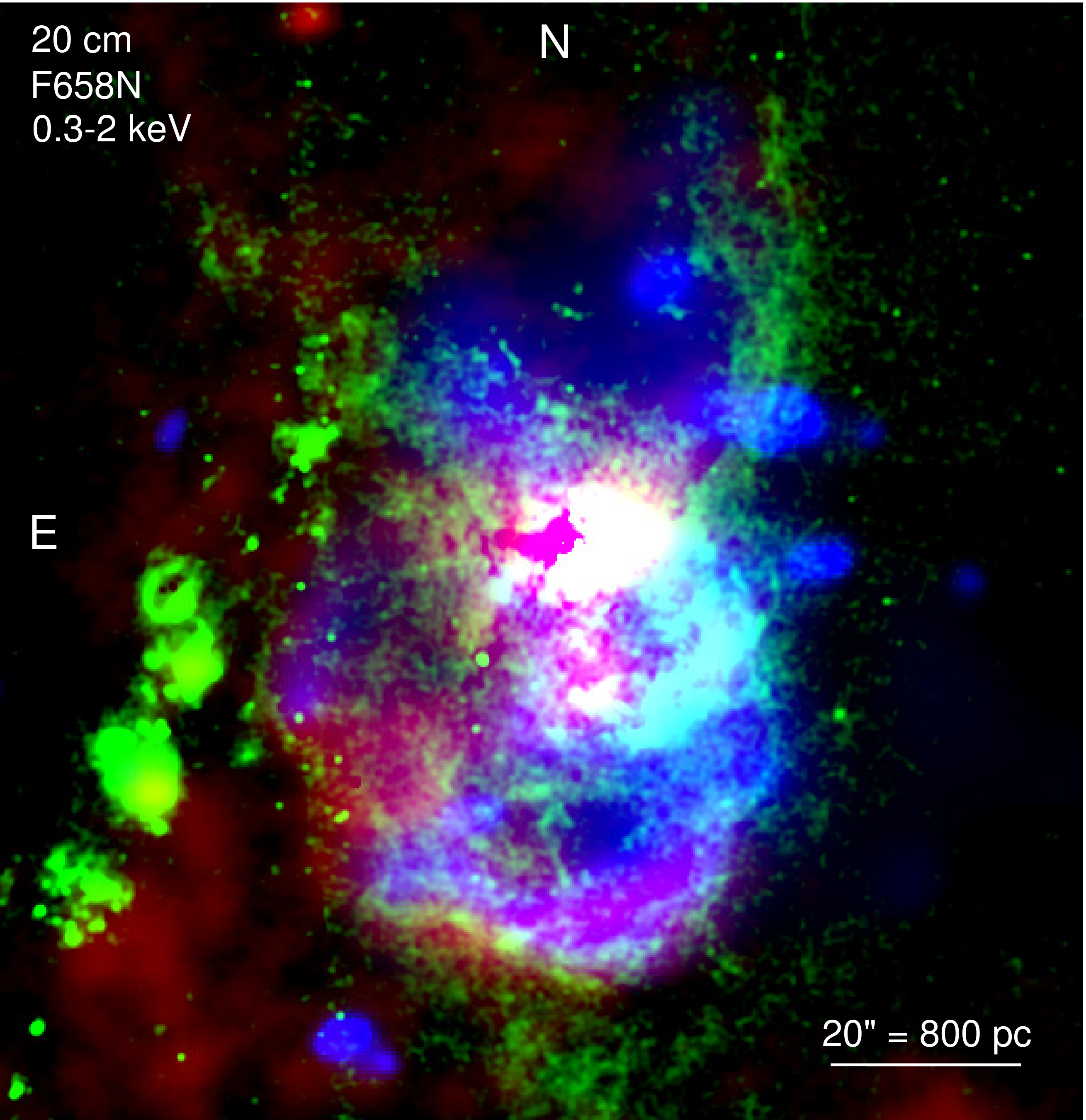}
\caption{Composite colour image in the radio band (red: VLA 20 cm), optical band (green: {\it HST}/ACS F658N filter) and X-ray band (blue: {\it Chandra}/ACIS-S 0.3--2 keV). Optical and X-ray images have been smoothed as in Figure 4. The scale bar is 20$''$ corresponding to $\approx$800 pc, and the size of the panel is $\approx$1$'$.9 $\times$ 2$'$0.}
\label{fig5new}
\end{figure}

The study by  \cite{SJMV} reported the detection of a pair of X-ray and H$\alpha$ arcs, one at $15''$ from the nucleus (the ``inner arc"), and another at $30''$ (the ``outer arc").  Figures \ref{fig4new} and \ref{fig5new} compare the \textit{Chandra}/ACIS-S image (0.3--2 keV), the \textit{HST}/ACS continuum-subtracted H$\alpha$ emission, and the VLA L-band map: there is considerable morphological similarity between the three bands. The radio spur appears to terminate in a region of bright X-ray and H$\alpha$ emission, but low radio emission. The radio arc appears morphologically similar to the X-ray and H$\alpha$ arcs, except for where the radio arc is connected to the radio spur.

\begin{figure*}
{\centering
\subfloat[Spectral index map]{\includegraphics[scale=0.5,clip=true,trim=3cm 1cm 0.5cm 1.5cm]{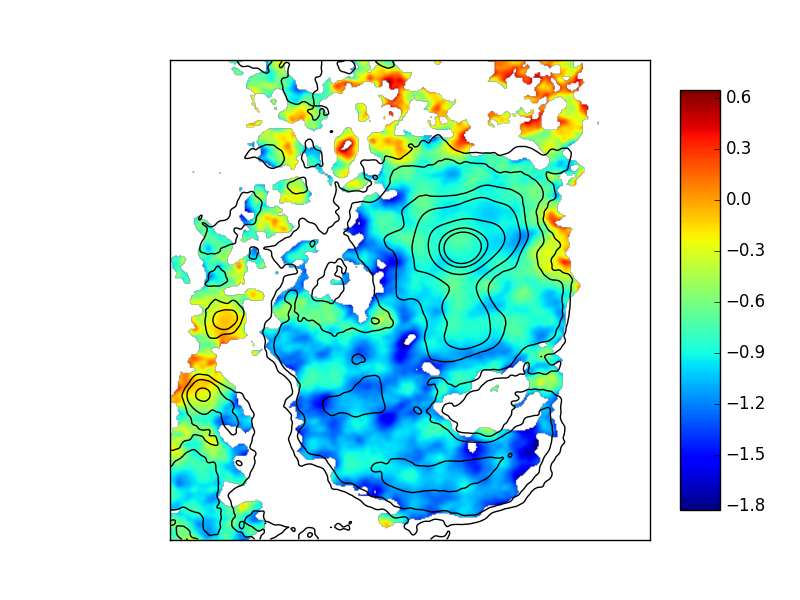}} \hspace{2em}
\subfloat[Error map]{\includegraphics[scale=0.5,clip=true,trim=3cm 1cm 0.5cm 1.5cm]{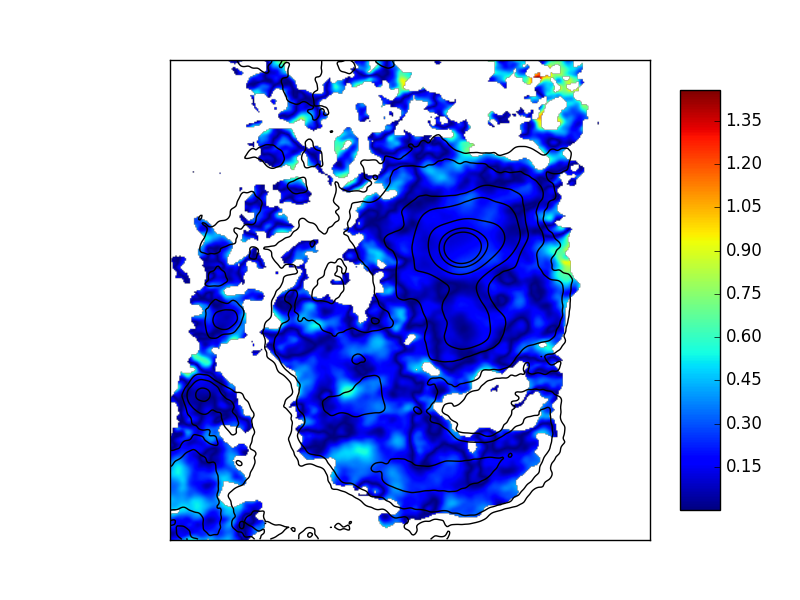}}

\caption{Spectral index map (left panel) and the corresponding error map (right panel), obtained from the combination of the VLA L-, C- and X-band images; L-band contours from \fig~\ref{fig:VLAMaps}a are overlaid. The image size is $1'.2 \times 1'.2$ ($\approx$ 2.8~kpc $\times$ 2.8~kpc). The spectral index gradually increases outwards from the core towards the radio arc.}
\label{fig:Specind}
}
\end{figure*}

To further understand the nature of the radio emission, we combined the VLA images in the L,  C, and X bands to create a spectral index map, using AstroPy and SciPy software (in particular, the sub-package {\it scipy.optimize.curve\_fit}). Since the VLA data have already been imaged using the same uvtaper, cellsize and weighting parameters, the C- and X-band images were re-gridded (via the {\small {AIPS}} task {\it regrd}) to match the L-band image. The three images were then exported as FITS files. 
To remove biasing towards steep $\alpha$ values due to noise and negative pixels, we imposed a 1$\sigma$ flux density cut on the three images, where $\sigma$ is the rms noise of each uvtapered image\footnote{We chose a 1$\sigma$ cut because the arc is detected at the 3--5$\sigma$ level in the L-band image but with lower significance at higher frequencies. Thus, the spectral index in the arc should be taken as an upper limit.}. The final spectral index and noise maps are shown in \fig~\ref{fig:Specind}. 

The spectral indices obtained for the core and spur ($-0.7 \la \alpha \la -0.6$) are consistent with optically thin, non-thermal synchrotron radio emission \citep{NWF01}, while the spectral index is significantly steeper within the arc ($\alpha < -0.7$); this is consistent with either jet activity or a population of supernova (SN) remnants  \citep{Mulcahyetal2016}. The lack of a self-absorbed flat-spectrum core ($\alpha \approx 0$) agrees with the non-detection of the radio core in the EVN data. The steepening of the spectral index in the arc could be the result of synchrotron ageing \citep{CPDL91}, or energy losses due to adiabatic expansion of the material as it does work on its surroundings \citep{CPDL91}. More observations
of the radio spectrum at lower frequencies are required for a more detailed analysis of either effect, but this is beyond the scope of this paper.

Finally, we compare our results with those from previous radio studies of NGC\,5195. The observations by \cite{Alataloetal2016} and \cite{NFW05} had worse angular resolutions than our images, but their reported fluxes are consistent with ours. \cite{Alataloetal2016} measured a total continuum flux density of $(1.21 \pm 0.21)$ mJy at 106 GHz with a beam size of $7''.6 \times 6''.1$, which is much larger than our VLA beam. \cite{NFW05} reported an upper limit of 1.1 mJy beam$^{-1}$ at 15 GHz, with a beam size of $0''.15$, close to the e-MERLIN $0''.2$ beam size at 1.5 GHz. If we assume an average spectral index $\alpha = -0.65$ in the core/spur region, we expect a flux density of 0.44 mJy at 15 GHz, consistent with the upper limit of \cite{NFW05}.
Our estimated spectral indices agree with those computed by \cite{Mulcahyetal2014}, who compared the flux densities at 151 MHz (from LOFAR) and at 1.4 GHz (from the VLA); they found $\alpha \approx -0.65$ for the nuclear region and $\alpha \approx -0.75$ for the outer arc.

\subsection{Balmer line emission}

We have already noted that the morphology of the optical line emission matches the location and extent of the radio and X-ray arcs, south of the nucleus. In addition, the Balmer line emission traces the contours of at least the lower half of what we suggested (Section 3.2) is the northern radio lobe or cocoon. In fact, the kpc-scale bipolar structure of the nuclear outflow is even clearer in the {\it HST} narrow-band image (Figures 3b and 3c) than in the VLA images. The bipolar structure (in particular, the lower edges of the northern lobe) opening up from the nuclear source is also detected in the {\it Chandra} image (Figure 3a), although at lower spatial resolution and signal-to-noise ratio. The existence of a similar bipolar structure at all three bands strongly suggests that most of the kinetic energy has been injected from a small region surrounding the nuclear position.

We used the {\it HST}/ACS image in the F658N filter, which includes H$\alpha$, [N {\footnotesize{II}}] $\lambda 6548$, and [N {\footnotesize{II}}] $\lambda 6583$. We defined three source extraction regions: a) a circle of radius 200 pc around the nuclear position, which covers most of the star-forming activity in the galaxy \citep{Alataloetal2016}; b) a 1-kpc-radius circular region covering what we shall call for simplicity the southern lobe (the inner and outer arcs described by SJVM, and the complex filamentary structure between and around them); c) a 1.2-kpc-radius circular region covering the northern lobe (the funnel-like structure opening up from the nucleus towards the north side of the galaxy, Figure~\ref{fig:intro}). We excluded the 200-pc nuclear region from both the northern and the southern lobe regions, so we can treat them separately. We also defined suitably large background regions outside the galaxy.
For each source region, we measured the background-subtracted count rates. We multiplied the count rates by the ACS flux-density normalization value PHOTFLAM\footnote{\url{https://acszeropoints.stsci.edu/}}, to convert count rates to flux densities; then we multiplied by the effective width of the filter, to obtain the combined flux in the lines covered by the filter. The effective width of the F658N filter is listed as 78 \AA\ in Table 5.1 of the ACS Instrument Handbook \footnote{\url{http://www.stsci.edu/hst/acs/documents/handbooks/current/c05_imaging2.html}}. We checked this value using the Python task {\small{PYSYNPHOT}}\footnote{\url{https://pysynphot.readthedocs.io/en/latest/index.html}} (as recommended by Jenna Ryon, ACS Instrument Team, priv.~comm.), under the plausible assumption that the three lines are much narrower than the filter width. We obtained an effective bandpass of 75 \AA\ with {\small{PYSYNPHOT}}, which is the value we adopt here. We also corrected the flux for line-of-sight Galactic extinction. The final result is that the total de-reddened flux in the F658N filter is: $F_{\rm{F658N}} = (1.9 \pm 0.4) \times 10^{-13}$ erg~s$^{-1}$~cm$^{-2}$ for the nuclear region; $F_{\rm{F658N}} = (5 \pm 1) \times 10^{-13}$ erg~s$^{-1}$~cm$^{-2}$ for the southern lobe; $F_{\rm{F658N}} = (3.5 \pm 0.5) \times 10^{-13}$ erg~s$^{-1}$~cm$^{-2}$ for the northern lobe.


The next step is to estimate what fraction of the flux in the F658N filter belongs to H$\alpha$, and what fraction belongs to the two [N {\footnotesize{II}}] lines. Generally, this ratio is a function of ISM density, metallicity, and shock velocity \citep{Allenetal2008}. Observations by \cite{Hoopes-Walterbos} suggest a ratio of [N {\footnotesize{II}}]$\lambda$6583/H$\alpha$ $\approx 1.3$ in the shock-ionized gas of the southern lobe (see their Table 5 and Fig.~19); we assume that the same ratio holds in the northern lobe. When the expected contribution from [N {\footnotesize{II}}]$\lambda$6548 is also taken into account (with a theoretically-predicted value of $F_{6583} \approx 2.95 F_{6548}$, almost independent of ISM conditions; \citealt{Allenetal2008}), we obtain $F_{{\rm{H}}\alpha} \approx
0.37F_{\rm{F658N}}$. For the star-formation-dominated nuclear region, we assume a ratio of [N {\footnotesize{II}}]$\lambda$6583/H$\alpha$ $\approx 0.3$ typical of photo-ionized H {\footnotesize II} regions \citep{Hoopes-Walterbos}, that is $F_{{\rm{H}}\alpha} \approx 0.7F_{\rm{F658N}}$. Finally, we assume a standard Balmer decrement for Case B recombination in warm gases ($T \approx 10^4$ K), and estimate $F_{{\rm{H}}\beta} \approx 0.35F_{{\rm{H}}\alpha}$. 
We obtain $F_{{\rm{H}}\beta} = (4.7 \pm 1.0) \times 10^{-14}$ erg~s$^{-1}$~cm$^{-2}$ in the inner 200 pc, 
$F_{{\rm{H}}\beta} = (6.5 \pm 1.3) \times 10^{-14}$ erg~s$^{-1}$~cm$^{-2}$ in the southern lobe, and $F_{{\rm{H}}\beta} = (4.5 \pm 0.5) \times 10^{-14}$ erg~s$^{-1}$~cm$^{-2}$ in the northern lobe. At the assumed distance of 8~Mpc, this corresponds to a luminosity $L_{\rm{H}\beta} = (5.0 \pm 1.0) \times 10^{38}$ erg~s$^{-1}$ in the southern lobe, $L_{\rm{H}\beta} = (3.4 \pm 0.5) \times 10^{38}$ erg~s$^{-1}$ in the northern lobe, and a total bubble luminosity $L_{\rm{H}\beta} = (8.4 \pm 1.1) \times 10^{38}$ erg~s$^{-1}$ (all values corrected for line-of-sight extinction).

For the nuclear region, the line-of-sight extinction is certainly an underestimate of the total extinction, which makes our estimate of the intrinsic Balmer line luminosity somewhat uncertain. If we take the column density estimated from the nuclear X-ray luminosity (Section 3.4), converted to an optical extinction with the relation of \cite{guver-ozel2009} ({\it i.e.}, $\approx$1 additional magnitude of intrinsic extinction in the V band), we need to multiply the observed H$\alpha$ luminosity by a factor of $\approx$2.
In summary, for the nuclear region we have an observed H$\alpha$ luminosity $L_{\rm{H}\alpha} \approx 10^{39}$ erg~s$^{-1}$, and we estimate an extinction-corrected intrinsic luminosity as high as $L_{\rm{H}\alpha} \approx 2 \times 10^{39}$ erg~s$^{-1}$.



\subsection{X-ray emission}
\begin{figure*}
\centering
\includegraphics[scale=0.25]{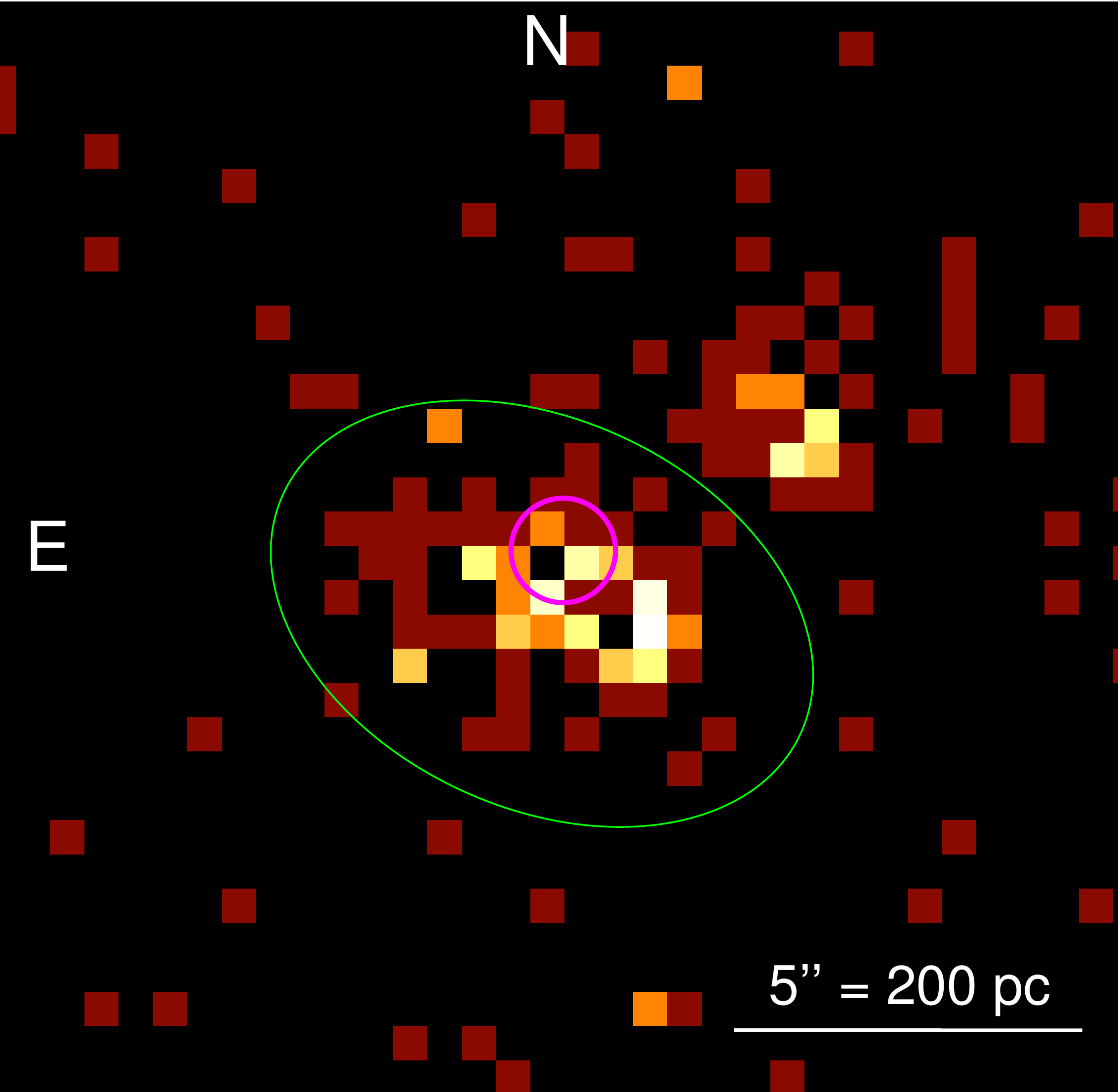}
\includegraphics[scale=0.25]{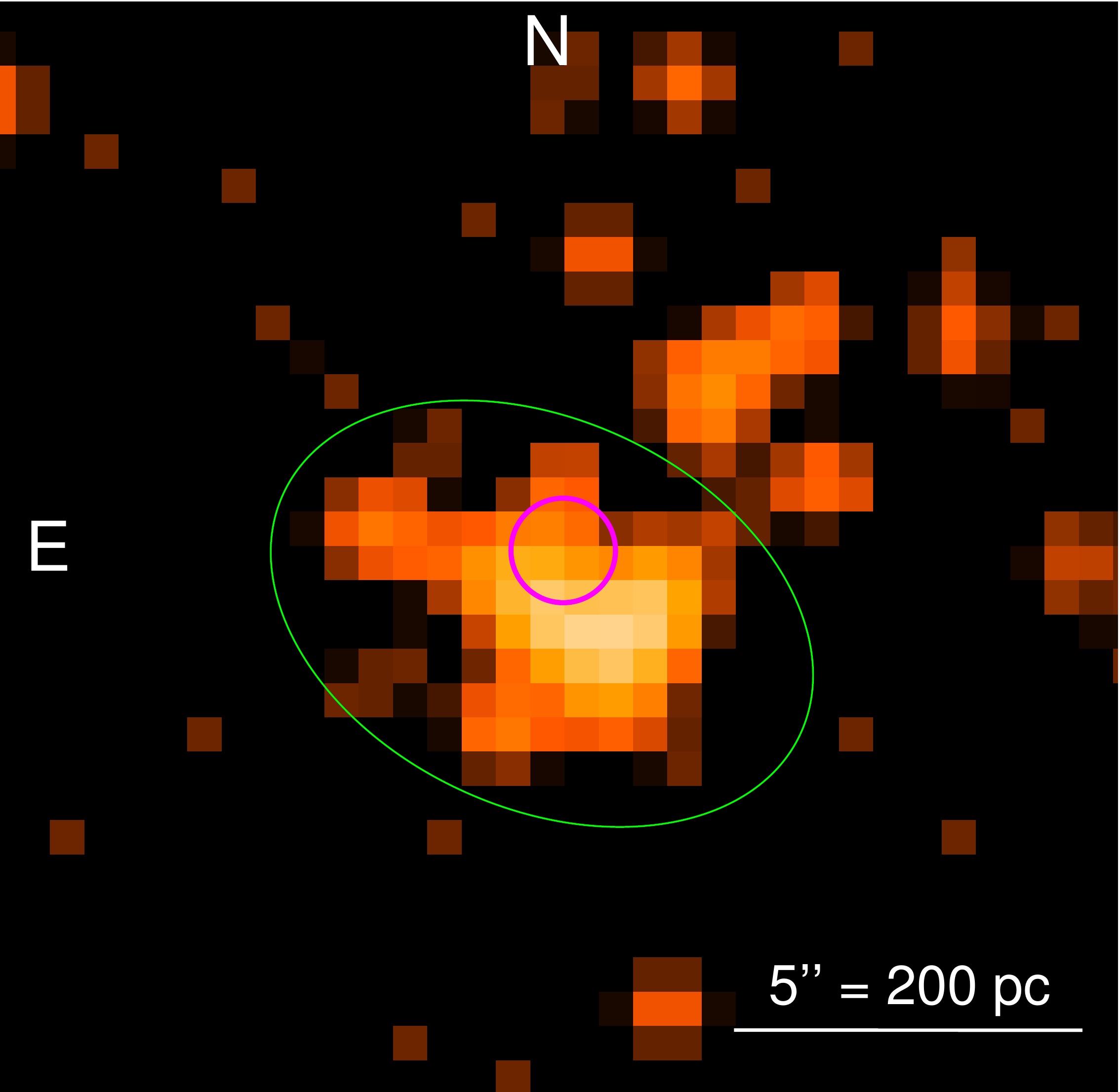}
\includegraphics[scale=0.25]{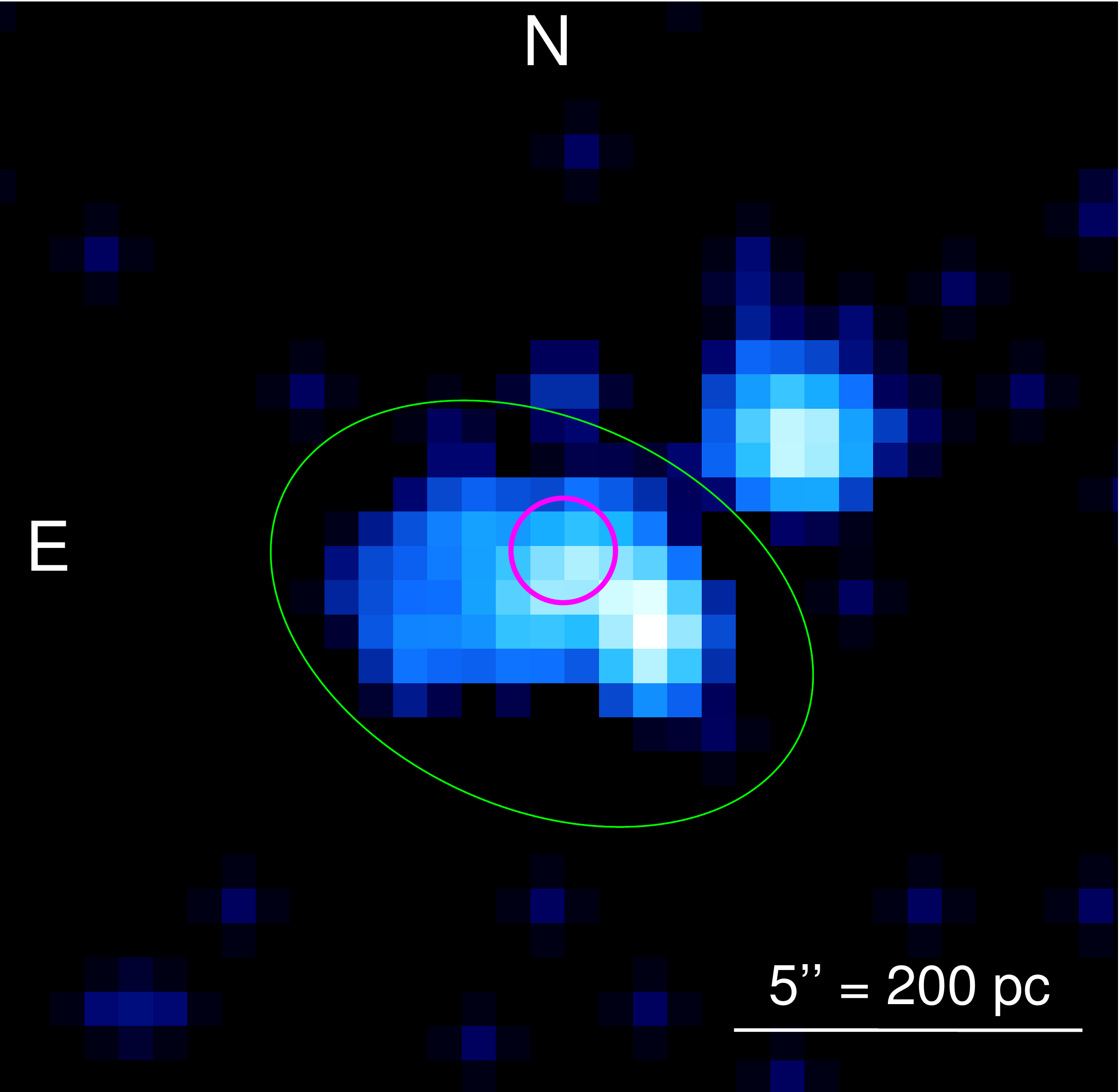}
\caption{Left panel: unsmoothed {\it Chandra}/ACIS-I image of the nuclear region, in the 0.3--7.0 keV band. Here and in the other two panels, the green ellipse is our source extraction region for the unresolved nuclear emission in the ACIS-S datasets; the magenta circle (radius of 0$''$.75) is the location of the radio core derived from our e-MERLIN observations. Middle panel: Gaussian-smoothed ACIS-I image, in the 0.3--1.5 keV band. Right panel: Gaussian-smoothed ACIS-I image, in the 1.5--7.0 keV band.}   
\label{acis-i}
\end{figure*}

The elongated nuclear emission seen in the stacked ACIS-S images is resolved into at least three sources in the ACIS-I image, with an additional point-like source located $\approx$4$''$ to the north-west of the nucleus (Figure~\ref{acis-i}). Based on the location of the radio core detected by e-MERLIN (Section 3.1), we find that only one of the partially resolved X-ray sources is consistent with the position of the radio nucleus (Figure~\ref{acis-i}). We used the emission in the 2--7 keV band to estimate the location of this compact X-ray source, in order to minimize the contamination from diffuse hot gas; we obtain RA $= 13^{\mathrm{h}}$29$^{\mathrm{m}}$59$^{\mathrm{s}}$.51, Dec $= +47^{\circ}$15$^{\prime}$58$^{\prime\prime}$.1 [J2000.0]. This is $\approx$0$''$.3 away from the e-MERLIN core position, well within the uncertainty of the absolute astrometry of the ACIS-I image (Section 2.5). Thus, this X-ray source is the most likely candidate for the supermassive BH emission. However, an X-ray binary origin also for this source cannot be completely ruled out by the available data.
Therefore, the luminosity of this source should be taken as the upper limit to the true X-ray luminosity of the supermassive BH.

Individual observations do not contain enough counts for meaningful spectral analysis; however, they do contain enough counts for a general assessment of long-term variability of the emission in the nuclear region. The sensitivity of the ACIS-S detector has degraded significantly over the years, especially below 1 keV; thus, before comparing count rates at different epochs, we converted the directly observed values to the nominal rates that would have been measured by the detector in 2012 (Chandra Cycle 13). Similarly, we converted the Cycle-18 ACIS-I count rate into an equivalent Cycle-13 ACIS-S count rate. For these conversions, we used the online tool {\small PIMMS}\footnote{\url{http://cxc.harvard.edu/toolkit/pimms.jsp}} version 4.8c, from the Chandra X-ray Center. We assumed a {\it mekal} model at $kT = 0.6$ keV to convert the 0.3--2.0 keV count rates, and a power-law model of index $\Gamma = 1.4$ for the 2--7 keV band: this choice approximates the average spectral properties illustrated later in this Section. (Using a power-law model for both bands does not significantly change the conversion.) We determined (Table~\ref{HR_tab}) that the net count rate (proxy for the X-ray luminosity) of the nuclear source varied by a factor of 3, with the observations at later epochs (2011--2012) fainter than those from 2000--2001. 
We do not have enough evidence to determine whether the decrease in luminosity is due to a dimming of the supermassive BH luminosity, or of one the other sources (likely X-ray binaries) in the nuclear region, unresolved by ACIS-S.


For our modelling of the stacked ACIS-S X-ray spectrum, we started from a simple power-law fit. We find a photon index $\Gamma \approx 1.63\pm0.10$, and intrinsic column density $n_{\rm H} = (0.13\pm0.03) \times 10^{22}$ cm$^{-2}$, in addition to a fixed Galactic line-of-sight column density of $1.8 \times 10^{20}$ cm$^{-2}$. However, this is a poor fit ($\chi^2_{\nu} = 275.8/147 = 1.88$), mostly because of a significant soft excess. We improve the fit with the addition of an optically thin thermal-plasma component (modelled with {\it mekal}: \citealt{Mewe1985, Mewe1986}, at the temperatures of $kT = 0.60^{+0.07}_{-0.09}$ keV. The best-fitting power-law photon index becomes $\Gamma \approx 1.43\pm0.10$, absorbed by a total column density $n_{\rm H} = 0.25^{+0.09}_{-0.06} \times 10^{22}$ cm$^{-2}$.  For the power-law plus thermal-plasma model, $\chi^2_{\nu} = 157.2/145 = 1.09$; an F-Test shows that this is a significant improvement over the simple power-law model, greater than the 99.9999\% level for the addition of the mekal component.

The unabsorbed luminosity of the power-law component in the 2--10 keV band is $L_{{\rm pl,}2-10} \approx (5.1 \pm 0.9) \times 10^{38}$ erg s$^{-1}$ at the assumed distance; in the 0.3--10 keV band, $L_{{\rm pl,}0.3-10} \approx (7.4 \pm 1.2) \times 10^{38}$ erg s$^{-1}$. Physically, the power-law component represents the contribution of compact accreting sources: supermassive BH and nearby X-ray binaries. The thermal plasma component is instead from diffuse hot gas, and its luminosity is a proxy for the nuclear star formation rate (SFR).  
For a direct comparison with the results of \cite{Mineoetal2012b} (see also Section 4.2), we fitted the {\it mekal} component with and without an intrinsic absorption term. When only Galactic line-of-sight absorption is included, the 0.5--2 keV thermal-plasma luminosity is 
$L_{{\rm{mek,}}0.5-2} \approx (0.7^{+0.3}_{-0.2}) \times 10^{38}$ erg s$^{-1}$; when a correction for the intrinsic absorption is included, the luminosity increases to $L_{{\rm{mek,}}0.5-2} \approx (1.5^{+0.8}_{-0.6}) \times 10^{38}$ erg s$^{-1}$, and about 10\% higher in the 0.3--10 keV band.

To estimate the current luminosity of the candidate supermassive BH, we measured its count rate from the 2017 ACIS-I image, and converted it to a flux by assuming that it has a power-law spectrum with the same parameters inferred from the stacked ACIS-S spectrum of the unresolved nuclear emission. We estimate an unabsorbed luminosity $L_{0.3-10} \approx (1.3 \pm 0.4) \times 10^{38}$ erg s$^{-1}$. The uncertainty is mostly due to the fact that the candidate nuclear source is only partly resolved, and its count rate may be contaminated by the nearby X-ray binaries and diffuse emission.


\begin{table*}
\caption{X-ray count rates and hardness ratios for the nuclear source$^{a}$}
\begin{tabular*}{\textwidth}{r @{\extracolsep{\fill}} cccc}
\hline\hline\\[-5pt]
ObsId & Obs Date & Rate$^{b}$ (0.5--2 keV) & Rate$^{b}$ (2--7 keV) & Hardness Ratio$^{b}$\\[2pt]
  &   &   ($10^{-3}$ s$^{-1}$) & ($10^{-3}$ s$^{-1}$) & \\[5pt]
\hline\hline\\[-5pt]
414 & 2000-01-23 & $7.5 \pm 2.6$ & $2.5 \pm 1.6$ & $0.33\pm0.24$\\
354 & 2000-06-20 & $4.8 \pm 0.6$ & $2.8 \pm 0.5$ & $0.60\pm0.12$\\
1622 & 2001-06-23 &  $5.3 \pm 0.4$ & $2.5 \pm 0.3$ &$0.47\pm0.07$\\
3932 & 2003-08-07 &  $4.6 \pm 0.3$ & $1.9 \pm 0.2$ &$0.41\pm0.05$\\
12668 & 2011-07-03 &  $2.8 \pm 0.6$ & $0.6 \pm 0.3$ &$0.23\pm0.13$\\
13813 & 2012-09-09 &  $2.91 \pm 0.14$ & $1.62 \pm 0.10$ &$0.56\pm0.04$\\
13812 & 2012-09-12  &  $3.42 \pm 0.16$ & $1.90 \pm 0.11$ &$0.56\pm0.04$\\
15496 & 2012-09-19 &  $1.75 \pm 0.22$ & $0.95 \pm 0.16$ &$0.54\pm 0.12$\\
13814 & 2012-09-20 &  $1.58 \pm 0.10$ & $1.19 \pm 0.08$ & $0.76\pm0.07$\\
19522 & 2017-03-17 &  $2.6 \pm 0.5$ & $1.5 \pm 0.3$ & $0.59\pm0.16$\\[5pt]
\hline\hline
\end{tabular*}
\begin{flushleft}
$^{a}$ Defined as the emission inside the green ellipse shown in Figure~\ref{acis-i}.\\
$^{b}$ Re-scaled to the equivalent values that would have been measured by ACIS-S in Cycle 13.
\end{flushleft}
\label{HR_tab}
\end{table*}


\section{Discussion}

We structure the Discussion into three main topics. First, in Section 4.1, we estimate the mechanical energy and long-term-average power required to generate the observed arcs and bubble. Then, in Section 4.2, we evaluate whether this power can be produced by multiple SNe associated with recent star formation, or instead requires an additional source of energy (AGN jet/outflows). Finally, in Section 4.3, using the constraints from our multi-band observations, we discuss whether the supermassive black hole is currently active, and whether we can distinguish between the emission contributions from the supermassive black hole and from a nuclear star-forming region.

\subsection{Energetics of the bubble} 
\label{sec:SF?}

\subsubsection{Kinetic power from bubble size and expansion speed}
To test the jet scenario for NGC\,5195, we need to look at the energetics of the bubble. It cannot have been inflated by a single or even a few explosive events, as its size and expansion speed require several orders of magnitude more energy than can be supplied for example by an individual SN or a tidal disruption event. Therefore, we shall assume a continuous injection of kinetic power, over a lifetime of a few Myr \citep{SJMV}.
This can be due either to a series of thousands of SNe in the nuclear region, or to an AGN jet. From standard bubble theory, in the Sedov-Taylor stage of expansion, the mechanical power required to inflate a bubble of radius $R$, expanding with a forward shock velocity $v_{\rm s}$ into an external medium of density $n$, is
\begin{equation}
P_{\rm kin} \approx 4.0 \times 10^{41} \, R_3^2 \, v_{\rm s,2}^3 \, n \ {\mathrm{erg~s}^{-1}}
\end{equation}
\citep{Pakulletal2006,Weaver+1977}, where $R_3 \equiv R/(1 {\rm \,kpc})$, and $v_{\rm s,2} \equiv v_{\rm s}/(100 {\rm {\,km~s}}^{-1})$. 

The expansion velocity and therefore the age of the bubble are not known with great accuracy; \citet{SJMV} suggest a range between 300--620  km s$^{-1}$, corresponding to ages between 5.6 Myr and 2.7 Myr, respectively. 
We estimate the undisturbed ISM density $n$ following the technique described by \citet{Pakulletal2006}, and \citet{Pakull+2010}. We compare our observed H$\beta$ luminosity of the bubble ($L_{{\rm{H}}\beta} \approx 8.4 \times 10^{38}$ erg s$^{-1}$: Section 3.3) with the predicted H$\beta$ flux emitted from the shocked region plus that from the precursor, that is the photo-ionized region ahead of the shock (equations 3.4 and 4.4 respectively, in \citealt{DopitaSutherland1996}). For fast shocks in the velocity range considered here, the H$\beta$ emission from the precursor is slightly higher (between 5\% and 15\%) than the emission from the shock itself \citep{DopitaSutherland1996}, hence it cannot be neglected. We then approximate the surface of the shock as a spherical bubble of radius $R$, and solve for the density $n$. We obtain:
\begin{equation}
n \approx (0.95 \pm 0.12) \, R_3^{-2} \, v_{\rm s,2}^{-2.41} \, \left(1+1.32 v_{\rm s,2}^{-0.13}\right)^{-1} \ {\rm cm}^{-3}
\end{equation}
which implies characteristic densities $n \sim 10^{-2}$ cm$^{-3}$, similar to those found in other early-type galaxies with jet bubbles \citep{Mingoetal2011}.

Finally, re-inserting the density into Equation 1, we obtain a kinetic power
\begin{equation}
P_{\rm kin} \approx (3.8 \pm 0.5) \times 10^{41}  \, v_{\rm s,2}^{0.59} \, \left(1+1.32 v_{\rm s,2}^{-0.13}\right)^{-1} \ {\mathrm{erg~s}^{-1}}
\end{equation}
For the likely range of shock velocities considered here, this corresponds to 
$P_{\rm kin} \approx (2.9$--$6.2) \times 10^{41}$ erg s$^{-1}$. 

\subsubsection{Kinetic power from the H$\beta$ emission line flux}
Let us now estimate the input mechanical power into the shock-heated gas directly from the flux of the H$\beta$ emission line \citep{DopitaSutherland1996,Pakull+2010}. For a fully-radiative shock expanding into the ISM, standard bubble theory \citep{Weaver+1977} predicts a total radiative luminosity $L_{\rm tot} \approx (27/77)P_{\rm kin}$ (the rest of the jet power going into the bulk kinetic energy of the expanding shell, and into the work done to expand into the ISM). Most of the photon luminosity is radiated in the form of IR/optical/UV lines. The fraction of luminosity emitted by any given line can be calculated using shock-ionization codes such as {\small MAPPINGS III} \citep{Allenetal2008}. The H$\beta$ line is particularly suited for this estimate because $L_{{\rm H}\beta}/L_{\rm tot}$ does not depend on the ISM density and has a relatively weak dependence on shock velocity \citep{DopitaSutherland1996}. For an equipartition magnetic field, we obtain that $L_{{\rm{H}}\beta}/P_{\rm kin} \approx 2.0 \times 10^{-3}$ for $v_{\rm s} = 300$ km s$^{-1}$, and $L_{{\rm{H}}\beta}/P_{\rm kin} \approx 1.1 \times 10^{-3}$ for $v_{\rm s} = 620$ km s$^{-1}$. This corresponds to a kinetic power
$P_{\rm kin} \approx (3.7$--$8.7) \times 10^{41}$ erg s$^{-1}$, in good agreement with our alternative estimate (Section 4.1.1) based on bubble size and expansion speed. Indeed, from our experience of jet bubbles around ultraluminous X-ray sources (ULXs; \citealt{FengSoria2011,Pakull+2010}), we find that the two methods usually agree within a factor of 2.
We conclude that (within the assumptions of standard bubble theory) the gas has been shocked by a long-term-average mechanical power of a few $10^{41}$ erg s$^{-1}$, corresponding to a total energy of a few 10$^{55}$ erg injected over its lifetime (2.7--5.6 Myr, from \citealt{SJMV}).

\subsection{Star-formation or nuclear jet?}

\subsubsection{Contribution from star formation}

Radio emission associated with star formation consists mainly of flat-spectrum free-free emission from H{\footnotesize {II}} regions and steep-spectrum synchrotron radiation from radio SNe and SN remnants \citep{Condon1992}. At frequencies below $\approx$30 GHz, we expect the synchrotron component to dominate. This is consistent with our spectral index map at centimetre wavelengths (Figure 6 and Section 3.2), which shows that the radio continuum from the arcs and bubble is non-thermal.

An SFR $\approx 0.14 M_{\odot}$ yr$^{-1}$ for the whole of NGC\,5195 (at an assumed distance of 7.7 Mpc) was derived by \citet{2013ApJ...768...90L} from their fitting of the spectral energy distribution from the UV ({\it GALEX}) to the far-infrared (FIR) ({\it Herschel}), although they also note that the FIR emission ``is confined to its nucleus". A re-analysis of this study by \citet{Alataloetal2016} suggested residual contamination to the FIR emission by NGC\,5194; thus, \citet{Alataloetal2016} revised the SFR of NGC\,5195 down to $\approx 0.12 M_{\odot}$ yr$^{-1}$, at their assumed distance of 9.9 Mpc. In our work, we have instead assumed a distance of 8.0 Mpc; this implies an SFR $\approx 0.08 M_{\odot}$ yr$^{-1}$. By comparing their FIR-derived SFR with the 1.4-GHz flux ($\approx$17 mJy) measured by the Westerbork SINGS survey \citep{2003PASP..115..928K, 2007A&A...461..455B} within the inner $\approx$800 pc, and using the radio-SFR relation of \citet{2011ApJ...737...67M, 2013ApJ...768....2M}, \citet{Alataloetal2016} found that the radio flux is only $\approx$20\% higher than expected from the SFR, and therefore they argue that the radio emission in NGC\,5195 ``is predominantly produced by star formation, rather than radio-loud AGN activity".

We can now compare the long-term-average mechanical power of the bubble (a few $10^{41}$ erg s$^{-1}$, as derived in Section 4.1), with the amount of kinetic power provided by the SFR quoted above.
We used Starburst99 \citep{Leitherer+1999,Leitherer+2014} to calculate the stellar plus SN kinetic power resulting from continuous star formation. Assuming a Kroupa initial mass function \citep{Kroupa01}, stellar evolutionary tracks from the Geneva code \citep{Ekstrom12}, and a metallicity Z$=0.014$, we obtain that $P_{\rm kin} \approx 4.0 \times 10^{41} \times$ SFR erg s$^{-1}$ (where SFR is in units of $M_{\odot}$ yr$^{-1}$). Thus, without assistance from an AGN jet, using only the stellar plus SN kinetic power, we need a SFR  $\approx 1 M_{\odot}$ yr$^{-1}$. This is an order of magnitude higher than the value of SFR $\approx 0.1 M_{\odot}$ yr$^{-1}$ derived earlier from the FIR luminosity \citep{Alataloetal2016}.

\subsubsection{Evidence in favour of an AGN jet contribution}
Our case for a recently active AGN jet rests on two observational results derived in our study. The first result is the morphology of the large-scale radio emission, and of the associated structures seen in H$\alpha$ and X-rays. We have shown that the radio emission has a peculiar spatial distribution, with arcs, a jet-like spur connecting the nucleus with the inner arc, and a large bipolar bubble, extended well beyond the region of current star formation. The second result is the energetics of the shock-ionized gas in the arcs and bubble, which requires an order of magnitude more power than available from current star formation, as discussed in Section 4.2.1; the SFR cannot have changed significantly over the lifetime of the bubble (a few Myr). 
Therefore, we agree with \cite{Alataloetal2016} that nuclear star formation plays a dominant role for the radio flux in the inner region ($\sim$100 pc). However, we argue that an additional AGN jet component is consistent with the data and may be responsible for the kpc-scale shock-ionized bubble/arc structure and for the apparent spur of radio emission extending southward of the nucleus in the VLA and e-MERLIN images. We suggest that the extended radio-continuum features revealed for the first time in this work are related to past epochs of AGN activity.



High resolution radio studies using e-MERLIN and VLBI have already shown that outflows and jets are common in nearby low-luminosity AGN ({\it e.g.}, \citealt{Krips+2007,GP09}); the peculiar structure of the radio, X-ray and optical emission in NGC\,5195 makes it an ideal case for studies of the large-scale effects of a jet on the ISM even in galaxies where the nuclear activity is currently very low. Kpc-scale bubbles with radio and X-ray emission (and in some cases also optical line emission) have been discovered around the nuclei of early-type galaxies such as Centaurus A \citep{Crostonetal2009,Kraftetal2003}, NGC\,3801 \citep{Crostonetal2007}, Markarian 6 \citep{Mingoetal2011}, as well as spiral galaxies such as NGC\,6764 \citep{Crostonetal2008}; typical jet powers in those galaxies are $\sim$10$^{41}$--10$^{42}$ erg s$^{-1}$. In some cases, starburst power and jet power may contribute a similar amount of power, as is thought to be the case in the spiral galaxy NGC\,3079 \citep{Shafi+2015,Cecil+2002}, which has a bubble morphology and energy very similar to what we find in NGC\,5195. At smaller scales, a kinetic power $\sim$10$^{39}$--10$^{40}$ erg s$^{-1}$ has been inferred for several super-Eddington ULXs \citep{Pakulletal2006,Pakull+2010}, which produce shock-ionized bubbles with a characteristic size of $\approx$100--300 pc.

Notice that so far we have estimated the possible AGN contribution from the indirectly inferred kinetic power of the bubble, rather than from the directly observed flux of the kpc-scale radio synchrotron emission. This is because non-thermal synchrotron emission is not a reliable proxy for jet power: the relation between kinetic power and synchrotron emission is poorly constrained and model dependent, function of (among other things) external density, magnetic field configuration, dissipation of energy along the jet path, and fraction of kinetic energy carried by protons and non-relativistic electrons \citep{GS2016, MH07, Birzan+2008,Cavagnolo+2010, BF2011}.
For low-power AGN in the sample of \citet{Cavagnolo+2010}, a jet power $\sim$10$^{41}$ erg s$^{-1}$
can correspond to 1.4-GHz synchrotron luminosities between $\sim$10$^{35}$ erg s$^{-1}$ and $\sim$10$^{38}$ erg s$^{-1}$. However, even this loose correlation has to be taken with great caution, because (as discussed by \citealt{GS2016}) the observed sample is affected by Malmquist bias \citep{Malmquist1922}, with both quantities correlating with the distances to the sources.

Alternatively, we compare the inferred power and observed radio luminosity of the AGN jet in NGC\,5195 with those estimated for large optical and radio bubbles around ULXs \citep{Pakulletal2002}. A kinetic power $\sim$10$^{40}$ erg s$^{-1}$ was inferred for some sources with non-thermal radio luminosities between $\approx$10$^{34}$ erg s$^{-1}$ and $\approx$2 $\times 10^{35}$ erg s$^{-1}$ \citep{2018MNRAS.tmp...14U}. Scaled up to a mechanical power of a few $10^{41}$ erg s$^{-1}$ (with $L_{1.4} \propto P_{\rm jet}^{0.7}$, typical of synchrotron emission), this corresponds to expected luminosities between $\sim$10$^{35}$ erg s$^{-1}$ and a few 10$^{36}$ erg s$^{-1}$. 

As we mentioned in Section 4.2.1, the 1.4-GHz flux density from the central region of NGC\,5195 reported by the SINGS survey \citep{2003PASP..115..928K, 2007A&A...461..455B}
is $\approx$17 mJy, corresponding to a luminosity $\approx$2 $\times 10^{36}$ erg s$^{-1}$; in our VLA observations (Table 3), we have split this component into a core contribution ($\approx$11 mJy) and a spur feature ($\approx$7 mJy). The total 1.4-GHz flux density (including the newly discovered outer arcs and bubble) measured in our study is $\approx$39 mJy, corresponding to $\approx$4 $\times 10^{36}$ erg s$^{-1}$ (Table 3). Thus, we suggest that about 30\% of the total synchrotron radio luminosity (mostly in the nuclear region) is due to star formation, and the rest (from the extended features) is the result of the interaction of the nuclear jet with the ISM. In summary, the jet-powered component of the radio luminosity is consistent with what we expect from the analogy of jet-driven ULX bubbles mentioned above.



\subsection{Current activity of the supermassive BH}
\label{sec:xrays_orign}

\subsubsection{Constraints from radio observations}
The EVN non-detection of a pc-scale radio core in 2011 suggests that a steady nuclear jet (if present) currently has very low power. For a flat spectrum, the flux density upper limit of 132 $\mu$Jy (Section 3.2) corresponds to a limit to its 8.6-GHz luminosity of $L_{\rm{8.6GHz}} < 8.7 \times 10^{34}$ erg s$^{-1}$. Using the conversion between radio core luminosity and jet power from \citet{Kordingetal2006a} (or the equivalent relation from \citealt{Heinzgrimm2005}), we infer a current limit to the jet power, $P_{\rm{jet}} \la 1 \times 10^{41}$ erg s$^{-1}$, and the mass accretion rate of the nuclear BH, $\dot{M} \la 10^{21}$ g s$^{-1}$. Using instead the alternative relation between observed radio luminosity and kinetic power in low-luminosity AGN, from \cite{MH07}, we obtain an even stronger upper limit $P_{\rm{jet}} \la 1 \times 10^{40}$ erg s$^{-1}$. Thus, the current jet power is at least a few times lower than the average value required to inflate the bubble over the last few Myr.  

At slightly larger scales, our e-MERLIN detection of a 0.3-mJy, optically thin compact core with an upper size limit of 1.3 pc is consistent with very recent nuclear jet activity. We suggest that the e-MERLIN optically-thin core is evidence that NGC\,5195 shows signs of episodic AGN activity, and harbours a recently re-ignited jetted AGN, with rapidly expanding, transient ejections rather than a self-absorbed steady jet. This scenario is similar to what has been found with VLBI observations in the ULX Holmberg II X-1, characterized by a spatially resolved, optically thin, adiabatically cooling core \citep{Cseh+2015} and possibly also in the other ULX IC\,342 X-1 \citep{Marlowe+2014}. It was suggested \citep{Cseh+2015} that the synchrotron radio and shock-ionized optical bubbles around some ULXs may be inflated by flaring ejections rather than a steady jet. We speculate that the same scenario may work in NGC\,5195, which shows a radio/optical/X-ray bubble remarkably similar to those found around several ULXs, although a factor of 10 larger in linear size. The estimates of the kinematic power presented in Section 4.1 only provide long-term-average values and cannot tell the difference between a series of flaring ejections and a steady jet. Flaring ejections are also more likely to lose collimation and transfer their kinetic energy to entrained ISM baryons closer to the nuclear region, consistent with the lack of hot spots in the arc region.

Moving further away from the nuclear BH, we have already noted that most of the optically thin radio emission detected by e-MERLIN and the VLA around the nuclear position on $\sim$10--100 pc scales is consistent with local star formation and with the FIR emission. However, the presence of an elongated feature in the e-MERLIN C-band image, roughly aligned with the north-south spur in the VLA images and with the major axis of the kpc bubble structure, may be a jet signature, and suggests that even at those intermediate scales, some of the radio emission is caused by recent episodes of increased jet activity.


\subsubsection{Constraints from X-ray observations}
In order to compare the estimated jet power with the observed X-ray luminosity, we need to determine the mass of the nuclear BH. We do so by using the $M_{\rm {BH}}$-$\sigma$ relation, knowing that the central velocity dispersion $\sigma = (127.4 \pm 4.9)$ km s$^{-1}$, from the Hyperleda database.
There are several alternative versions of the $M$-$\sigma$ relation in the published literature, leading to slightly different estimates. From the best-fitting relation to the sample of 21 barred galaxies in \citet{Grahamscott2013}, we estimate $\log (M_{\rm {BH}}/M_{\odot}) = 6.9$ with a scatter of 0.4 dex; from the fit to the whole sample of 72 galaxies in the same paper \citep{Grahamscott2013}, we obtain $\log (M_{\rm {BH}}/M_{\odot}) = 7.0$ with the same scatter. From the $M_{\rm {BH}}$-$\sigma$ relation in \citet{vandenBosch2016}, we obtain instead $\log (M_{\rm {BH}}/M_{\odot}) = 7.3$ with a scatter of 0.5 dex. Given the scatter in the relation, we adopt a mass $M_{\rm{BH}} = 10^7 M_{\odot}$ for the rest of this Section. By inserting the BH mass and the current upper limit to the 5-GHz core radio luminosity into the fundamental plane relation of \citet{Plotkinetal2012}, we obtain an upper limit to the 0.5--10 keV luminosity of the AGN: $L_{0.5-10} \la 1.3 \times 10^{38}$ erg s$^{-1}$ $\approx 10^{-7} L_{\rm {Edd}}$. 
This value is similar to the observed X-ray luminosity of the (partially resolved) source closest to the position of the e-MERLIN radio nucleus ($L_{0.5-10} \approx 1.2 \times 10^{38}$ erg s$^{-1}$; Section 3.4). Therefore, we conclude that the observed radio and X-ray luminosities of the supermassive BH are self-consistent, and confirm that the current level of AGN activity is very low.

As discussed in Section 3.4, the nuclear X-ray emission within $\approx$200 pc contains also a thermal-plasma component with $L_{{\rm mek},0.3-10} \approx 1.6 \times 10^{38}$ erg s$^{-1}$, after accounting for intrinsic absorption. By comparing with the $L_{\rm X}$-SFR relation of \citet{Mineoetal2012a}, we infer that such luminosity is expected for an SFR of $\approx$0.02 $M_{\odot}$ yr$^{-1}$. Analogous relations between SFR and diffuse hot gas luminosity were also extensively studied by \citep{2018AJ....155...81S} for a sample of merging or interacting galaxies: from their results, we confirm that the observed thermal plasma luminosity in the nucleus of NGC\,5195 requires an SFR of $\approx$0.02 $M_{\odot}$ yr$^{-1}$. Several other studies of the relation between diffuse X-ray luminosity and SFR \citep{Lehmeretal2010, Kaaret08,Persic07,Ranalli03} confirm this estimate, with mutual discrepancies within a factor of 2. The SFR inferred for NGC\,5195 is $\approx 0.08 M_{\odot}$ yr$^{-1}$ (as discussed in Section 4.2.1) but a more constraining upper limit for the nuclear SFR is $\approx$0.05 $M_{\odot}$ yr$^{-1}$ estimated within $\approx$800 pc (\citealt{Alataloetal2016}, rescaled to our assumed distance). From our HST measurement of the intrinsic H$\alpha$ luminosity in the inner 200 pc (Section 3.3), using the standard conversion between H$\alpha$ luminosity and SFR \citep{kennicutt1998}, we estimate a SFR $\approx$0.01--0.02 $M_{\odot}$ yr$^{-1}$ in that small region, in agreement with the thermal-plasma X-ray luminosity.

\subsubsection{Constraints from high-ionization lines}
As an aside, we can re-examine the detection of a [Ne {\footnotesize{V}}] line at 14.32 $\mu$m, in the nuclear region, with a flux of $(2.0 \pm 0.6) \times 10^{-15}$ erg cm$^{-2}$ s$^{-1}$ \citep{GA2009}; this corresponds to a luminosity of $\approx$1.5 $\times 10^{37}$ erg s$^{-1}$, about 10\% of the H$\beta$ luminosity from the same region. If the line is entirely due to X-ray photo-ionization, the empirical relations between the [Ne {\footnotesize{V}}] flux, the [O {\footnotesize{III}}] flux, and the continuum luminosity in AGN \citep{Lamastra+2009,GA2009} suggest that a 2--10 keV luminosity $L_{\rm X} \approx 5 \times 10^{39}$ erg s$^{-1}$ is required to ionize the gas: an order of magnitude higher than the current nuclear X-ray luminosity. If the [Ne {\footnotesize{V}}] line is due to shock-ionization (including the shock precursor flux), it points to shock velocities $\ga$500 km s$^{-1}$ in the nuclear region \citep{Allenetal2008}. A more active nuclear BH in recent epochs (over a recombination timescale) and fast shocks from a nuclear jet are both consistent with the jet bubble scenario outlined in our paper.


\section{Summary and Conclusions}

The interacting galaxy NGC\,5195 has undoubtedly seen a lot of energy injected into its ISM from the central region. As previously reported \citep{Alataloetal2016}, the radio emission from the nuclear region within $\sim$100 pc is non-thermal synchrotron emission with an integrated flux that is marginally consistent with (although slightly higher than) what is expected from star formation alone. However, in this work we have identified extended low-surface brightness features in the radio-continuum, which point to previous episodes of jetted AGN activity. We have mapped such features by combining EVN data at pc-scale resolution, e-MERLIN data at $\sim$10-pc resolution, and VLA data at $\sim$100--1000 pc scales. We have shown that such radio structures are associated with arcs of X-ray emitting plasma, and with filaments of shocked gas (traced by H$\alpha$ line emission). 

The nuclear region is surrounded by a kpc-scale bubble-like structure, with a bipolar shape; it extends about 1.7 kpc on the southern side, and more than 3 kpc (in the radio) on the northern side. Emission features are stronger on the southern side of the bubble, possibly because of the higher ISM density towards the much bigger companion galaxy NGC\,5194; however, the outlines of the northern side of the bubble are also clearly seen in X-rays, optical and radio. The bubble size and structure favour the jet over the starburst interpretation. The bipolar shape of several structures (in particular the H$\alpha$ emission) is consistent with an outflow expanding preferentially along a north-south direction, rather than in a spherically symmetric way. If at the origin of this baryonic outflow there is a nuclear jet, it is likely that it interacts with the surrounding medium and transfers kinetic energy already close to the nucleus, as is the case in low-power AGN. From the size and expansion velocity of the kpc-scale bubble, and from the Balmer line emission of the recombining gas, we estimated an average kinetic power $\approx$3--6 $\times 10^{41}$ erg s$^{-1}$, injected over the last few Myr.

We did not detect a flat-spectrum core at VLBI scale (beam size 0.38 pc $\times$ 0.22 pc), down to an upper limit of 0.13 mJy (EVN observation taken in 2011 November). This is an order of magnitude lower than the long-term-average power inferred from the extended radio/optical/X-ray structures. One scenario is that the nuclear BH does not have a steady jet (identified by a self-absorbed flat-spectrum core), and is instead characterized by off states and periodic episodes of jetted AGN activity. The current low radio luminosity of the supermassive BH is self-consistent (via fundamental-plane relations) with its low X-ray luminosity ($\approx$10$^{38}$ erg s$^{-1}$), inferred from {\it Chandra} observations. Most of the X-ray emission in the inner $\approx$100 pc is due to a few X-ray binaries and diffuse hot gas; in particular, the observed luminosity of the thermal plasma component in the nuclear region is consistent with the SFR inferred from the FIR and H$\alpha$ measurements.

In conclusion, we suggest that NGC\,5195 is a very interesting test case for physical models of moderately powerful AGN jets, such that most of their kinetic energy is transferred to the circumnuclear ISM within a kpc scale, forming arcs, bubbles and filaments of shocked gas in the ambient medium. It is also a test case for the timescale of jet variability, and for the duty cycle of AGN phases in normal galaxies.


\section*{Acknowledgements}

We thank the anonymous referee for their detailed comments and suggestions and their patience through various earlier versions of this work, which was much improved as a result. We also thank D.~Mulcahy and M.~W.~Pakull for invaluable discussions. 
e-MERLIN is a National Facility operated by the University of Manchester at Jodrell Bank Observatory on behalf of STFC. The VLA is operated by the NRAO is a facility of the National Science Foundation operated under cooperative agreement by Associated Universities, Inc. The Hubble Space Telescope, is operated by NASA and the Space Telescope Science Institute. The Chandra X-ray telescope is operated for NASA by the Smithsonian Astrophysical Observatory. 
This work is partly based on observations made with the NASA/ESA Hubble Space Telescope, and obtained from the Hubble Legacy Archive, which is a collaboration between the Space Telescope Science Institute (STScI/NASA), the Space Telescope European Coordinating Facility (ST-ECF/ESA) and the Canadian Astronomy Data Centre (CADC/NRC/CSA). 
We used material from the NASA/IPAC Extragalactic Database (NED), operated by the Jet Propulsion Laboratory, California Institute of Technology, under contract with NASA. We also used the VizieR catalogue access tool, and APLpy \citep{RB2012}, an open-source plotting package for Python. HR acknowledges support from ERC-StG 307215 (LODESTONE). RS acknowledges support from a Curtin University Senior Research Fellowship; he is also grateful for support, discussions and hospitality at the Strasbourg Observatory during part of this work. RU acknowledges that this research is supported by an Australian Government Research Training Program (RTP) Scholarship. The International Centre for Radio Astronomy Research is a joint venture between Curtin University and the University of Western Australia, funded by the state government of Western Australia and the joint venture partners.

\small
\bibliographystyle{mnras}
\bibliography{M51b}

\end{document}